\documentclass[traditabstract]{aa} 
\usepackage{multirow}

\newcommand{\Ms}{M$_{\odot}$}

\newcommand{\al}{Al$_2$O$_3$}

\newcommand{\mic}{$\mu$m}

\newcommand{\kms}{km s$^{-1}$}
\newcommand{\cmc}{cm$^{-3}$}

\usepackage{graphicx}
\usepackage{amsmath}

\usepackage{txfonts}
\begin{document}

   \title{Molecules and dust in Cassiopeia~A: II - Dust sputtering and diagnosis of supernova dust survival in remnants.  }

   \author{Chiara Biscaro\inst{\ref{inst1},}\inst{\ref{inst2}} \& Isabelle Cherchneff\inst{\ref{inst2}}
}

   \institute{Dark Cosmology Centre, University of Copenhagen, Juliane Maries Vej 30, 2100 Copenhagen, Denmark\label{inst1} \\
   \email{chiara@dark-cosmology.dk}
  \and
Physik Departement, Universit{\"a}t Basel, Klingelbergstrasse 82, 4056 Basel, Switzerland\label{inst2}\\
\email{isabelle.cherchneff@unibas.ch}}
   \date{}

 
  \abstract{
   We study the dust evolution in the supernova remnant Cassiopeia A. We follow the processing of dust grains that formed in the Type~II-b supernova ejecta by modelling the sputtering of grains. The dust is located in dense ejecta clumps that are crossed by the reverse shock. We also investigate further sputtering in the inter-clump medium gas once the clumps have been disrupted by the reverse shock. The dust evolution in the dense ejecta clumps of Type II-P supernovae and their remnants is also explored.  
   
  We study oxygen-rich clumps that describe the oxygen core of the ejecta, and carbon-rich clumps that correspond to the outermost carbon-rich ejecta zone. We consider the various dust components that form in the supernova, several reverse shock velocities and inter-clump gas temperatures, and derive grain-size distributions and masses for the dust as a function of time. Both non-thermal sputtering within clumps and thermal sputtering in the inter-clump medium gas are studied.
    
   We find that non-thermal sputtering in the clumps is important for all supernova types and accounts for reducing the grain population by $\sim 40$ \% to $80$~\%\ in mass, depending on the clump gas over-density, the grain type and size, and the shock velocity in the clump. A Type II-b SN forms small grains that are sputtered within the clumps and in the inter-clump medium. For Cas A, silicate grains do not survive thermal sputtering in the inter-clump medium, while alumina, silicon carbide, and carbon dust may survive in the remnant. Our derived masses of currently processed silicate, alumina and carbon grains agree well with the values derived from the observations of warm dust, and seem to indicate that the dust is currently being processed within clumps by non-thermal sputtering. Out of the $\sim 0.03$~\Ms\ of dust formed in the ejecta, between 30\% and 60 \% of this mass is present today in Cas A, and only 6 \% to 11 \% of the initial mass will survive the remnant phase. Grains formed in Type II-P supernovae are larger and better survive their journey in the remnant with mass fractions of surviving dust in the range $14-45$ \%. For very dense ejecta, which describe the ejected material in supernovae such as SN1987A, non-thermal sputtering is inefficient within the clumps and dust survival efficiencies in the remnant range between 42\% to 98\%\ in mass. For the SN1987A model, the derived surviving dust mass is in the range $\sim 0.06-0.13$~\Ms. This type of supernovae with dense ejecta may then efficiently provide galaxies with dust. Specifically, silicate grains over 0.1 \mic\ and other grains over $0,05$~\mic\ survive thermal sputtering in the remnant. Therefore, pre-solar grains of supernova origin that are found in meteorites, in particular silicate and alumina grains, possibly form in the dense ejecta clumps of Type II-P supernovae.}

   \keywords{ISM: Supernova remnant; dust, extinction; Supernovae: general, Circumstellar matter }
   
      \titlerunning{Dust sputtering in the supernova remnant Cas~A}
   
     \authorrunning{Biscaro \& Cherchneff}
     
       \maketitle

\section{Introduction}

Evolved massive stars with M$_{\star} \ge 8$ \Ms\ explode as core-collapse supernovae (SNe). The most common SNe are of Type~II-P, with a stellar progenitor mass on the ZAMS in the range $\sim 9-25$~\Ms\ (\cite{ham03,sma09}). A  year after outburst, evidence for the production of molecules and dust grains in the hot ejecta of several Type~II~SNe was provided by infrared (IR) observations (e.g. Lucy et al. 1989, Wooden et al. 1993, Kotak et al. 2009). The dust then travels behind the forward explosion blast wave within the ejecta and gets reprocessed during the SN remnant (SNR) phase. SNRs are characterised by the formation of a reverse shock (RS) when the high pressure material, which  is swept-up by the forward shock, expands and pushes back on the stellar ejecta (\cite{che77}). The RS crosses back the ejecta and reprocesses any existing ejecta material, i.e. molecules and dust grains. Therefore, the net budget of dust produced by SNe must account for the total amount of dust that is formed in the nebular phase and that survives processing by the RS in the SNR phase. 

Recent submillimetre (submm) observations with the space telescopes {\it AKARI} and {\it Herschel}, and the ground-based interferometer array ALMA seem to indicate the presence of large masses of dust grains in a few SNRs, specifically SN1987A (Matsuura 2011, 2015a; Indebetouw et al. 2014), Cas~A (\cite{sib10,bar10} and the Crab~Nebula (\cite{gom12,tem13}), with values in the range $0.02-0.8$~\Ms. These values are higher than the dust masses derived from IR observations in SNe by several orders of magnitude, and the origin of this discrepancy has been highly debated (Sarangi \& Cherchneff 2013, 2015, Wesson et al. 2015, Dwek \& Arendt 2015). The assessment of the net dust mass an SN+SNR system produces is  important in explaining the large quantities of dust inferred from the reddening of Ly$-\alpha$ systems and background quasars at high redshift, e.g. J1148+5251 at $z=6.4$ (\cite{pei91, pet94, hi06}). Indeed, the explosion of massive stars have played a pivotal role in the enrichment of primitive galaxies (\cite{alva06}). 

Extensive modelling efforts have been undertaken to explain observations and the formation of dust in local and high redshift SNe. While some studies use the classical nucleation theory (CNT) to describe dust formation under equilibrium conditions in shocked environments (Kozasa et al. 1989, Todini \& Ferrara 2001, Nozawa et al. 2003, Schneider et al. 2004), other studies describe the formation of both molecules and dust clusters from the shocked gas by assuming a chemical kinetic approach under non-equilibrium conditions and the subsequent coalescence and coagulation of these clusters to form dust grains (Cherchneff \& Lilly 2008, Cherchneff \& Dwek 2009, 2010, Sarangi \& Cherchneff 2013, 2015, hereafter SC15). On the other hand, the processing of dust in SNRs has received less attention. The thermal sputtering of dust by the forward shock has been studied by Nozawa et al. (2006), and sputtering by the RS in local and primitive SNe was modelled by Nozawa et al. (2007), Bianchi \& Schneider (2007), and Nath et al. (2008). Sputtering by the RS in Type~II-b SNe with application to the Cas~A SNR was studied by Nozawa et al. (2010). Finally, Silvia et al. (2010, 2012) proposed a hydrodynamic study of an ejecta clump, which was processed by the RS, and followed thermal sputtering as the clump was crossed and gradually disrupted by the shock. All studies show that dust is strongly reprocessed in SNRs. However, the final mass of surviving dust is not well constrained because of several factors. Firstly, all studies use as initial conditions for sputtering in the SNR the dust masses and size distributions that result from applying CNT. The chemical types and masses of the dust that are used as initial conditions may consequently be in error using this formalism, as discussed by Cherchneff (2014). Furthermore, some of these studies assume a homogeneous ejecta in the SNR phase (Nozawa et al. 2007, 2010; Bianchi \& Schneider 2007; Nath 2008), which implies that very harsh conditions hold in the post-shock gas for dust survival because the RS crosses and reprocesses the ejecta at very high velocities. By considering clumpy ejecta, the RS velocity will be greatly reduced in the dense clumps (\cite{sil10,bis14}). 

Indirect evidence for SN dust survival is proposed  by the studies of meteorites (\cite{zin07}). Some pre-solar grains bear the isotopic anomalies typical of Type~II-P SN nucleosynthesis. Thus, the dust that is formed in SN ejecta has survived the SNR phase and its journey in the interstellar medium, where other shock processing occurs, to finally enter the solar system. These pre-solar grains are of different chemical types, namely alumina, \al, silicates, silicon carbide, SiC, and carbon, and are usually large, with a size in the range $0.1-1$ \mic\ (\cite{hop11}). Other pre-solar grains of SN origin have a more exotic composition, such as silicon nitride, Si$_3$N$_4$, and quartz, SiO$_2$ (\cite{zin07, haen13}). These pre-solar grains seem to indicate possible mild processing conditions in the SNR phase. Such conditions exist when the RS reprocesses a clumpy ejecta. 

The 340~year-old SNR Cas~A has been extensively studied through observations at wavelengths spanning the radio to the X-ray domain. Optical photometric data indicate the SNR material is knotted and clumpy (Fesen et al. 2006, 2011) and the 3D reconstructions of optical emission show  complex remnant dynamics, including jets and rings of shocked ejecta material (\cite{del10, ise12, mil13}). Evidence for warm dust and CO emission was proposed from IR photometry observations with {\it Spitzer} (\cite{rho08, rho09}); the RS currently crosses the ejecta towards the collapsed star and reprocesses the ejecta material. The detection of warm dust relates to the ejecta grains that are shocked and sputtered by the RS. The recent detection of high excitation lines of warm CO with Herschel, possibly originating from the post-RS warm gas in a CO-rich clump already identified with {\it Spitzer}, supports the presence of shocked ejecta gas in Cas~A (\cite{wal13}). 

The fate of molecules and dust clusters in an oxygen-rich ejecta clump that has been shocked by the RS in Cas A has been studied by Biscaro \& Cherchneff (2014, hereafter BC14). They consider clumps, which are denser than a homogeneous ejecta by factors of 200 to 2 000 and pre-shock gas densities of 100~\cmc, which are derived from the far-infrared line analysis of Docenko \& Sunyaev (2010). Conditions in the clump's post-shock gas are taken from the model of radiative shocks in fast-moving oxygen-rich knots in Cas A by Borkowski \& Shull (1990). Post-shock gas densities in the range $10^5-10^6$~\cmc\ are derived and are in good agreement with density values inferred from the observations of high excitation CO lines in a dense clump in Cas~A by Walstr{\"o}m et al. (2013). The BC14 study highlights the importance of considering over-densities when modelling the formation of dust grains in Type II-b SNe by using a non-equilibrium chemistry and a chemical kinetics approach. Indeed, for the homogenous gas conditions given by the explosion models of Type II-b SNe (e.g., Nozawa et al. 2010), only very small amounts of dust clusters and grains can form and the variety of dust types is poor. In dense SN clumps, the formation of dust is efficient and reasonable dust masses are obtained with a large range of dust chemical types. As shown by BC14, ejecta molecules are destroyed by the RS but reform, specifically CO, in the post-shock gas in the clump. However, dust clusters cannot form out of the gas phase in the post-shock gas. Therefore, any dust destroyed by the RS is lost to the total dust budget that is produced by a SN+SNR system. 

Following this study on the molecular component of RS-shocked ejecta clumps in Cas A, we now investigate the non-thermal sputtering of dust in similar clumps that are shocked by the RS, and follow the dust thermal sputtering in the hot inter-clump ejecta gas once the clump has been totally disrupted by the RS. We consider several shock velocities and dust types that are typical of oxygen-rich clumps (silicates and alumina) and carbon-rich clumps (carbon and silicon carbide). The grain masses and size distributions are derived from modelling dust formation in the ejecta of a Type~II-b SN, in the case of Cas~A, and of Type~II-P SNe (SC15). Several inter-clump gas temperatures, including those typical of the Cas~A gas, are considered. The clump model is presented in Section~\ref{init}, and the sputtering model is explained in Section~\ref{sput}. The results on dust sputtering are presented for Cas~A  and the SNR resulting from a Type~II-P SN, and are compared to other studies on sputtering in Section~\ref{res}. A summary and discussion follow in Section~\ref{dis}.

\section{Chemical and physical model}
\label{init}
The importance of using a kinetic approach to describe the ejecta chemistry of the SN that led to the Cas~A remnant and the nucleation of dust clusters is discussed in BC14, where the effect of the ejecta gas density on the formation of complex molecules and dust clusters is studied. Specifically, the Cas~A SN ejecta has to be clumpy to form dust clusters of various chemical types and in large enough quantities. We  find that these clumps must be between 200 and 2 000 times denser than the initial Type~II-b homogeneous ejecta considered by Nozawa et al. (2010) to form sufficient dust masses of various chemical compositions. While an over-density factor of 200 is appropriate for the Cas A clumpy gas, a factor of 2 000 results in gas densities typical of the homogeneous ejecta of Type~II-P~SNe (Sarangi \& Cherchneff 2013, SC15). 

 First of all, we need to calculate the grain size distributions and masses that are produced by the Cas~A SN progenitor to study dust sputtering in the remnant phase. Therefore, we use the results on dust cluster masses and types derived in BC14 and apply the model developed by SC15 to describe the condensation of these dust clusters into grains. The study considers that the dust cluster seeds form out of the gas phase by pure chemical processes, and grow by coalescence and coagulation to produce dust grains. The formalism is based on a Brownian diffusion description (\cite{jac05,sp06}), and we direct the reader to SC15 for an exhaustive description of the formalism and method.

\subsection{Dust masses and grain size distributions from the Cas~A supernova}
\label{GSD}

\begin{figure}
\centering
\includegraphics[width=\columnwidth]{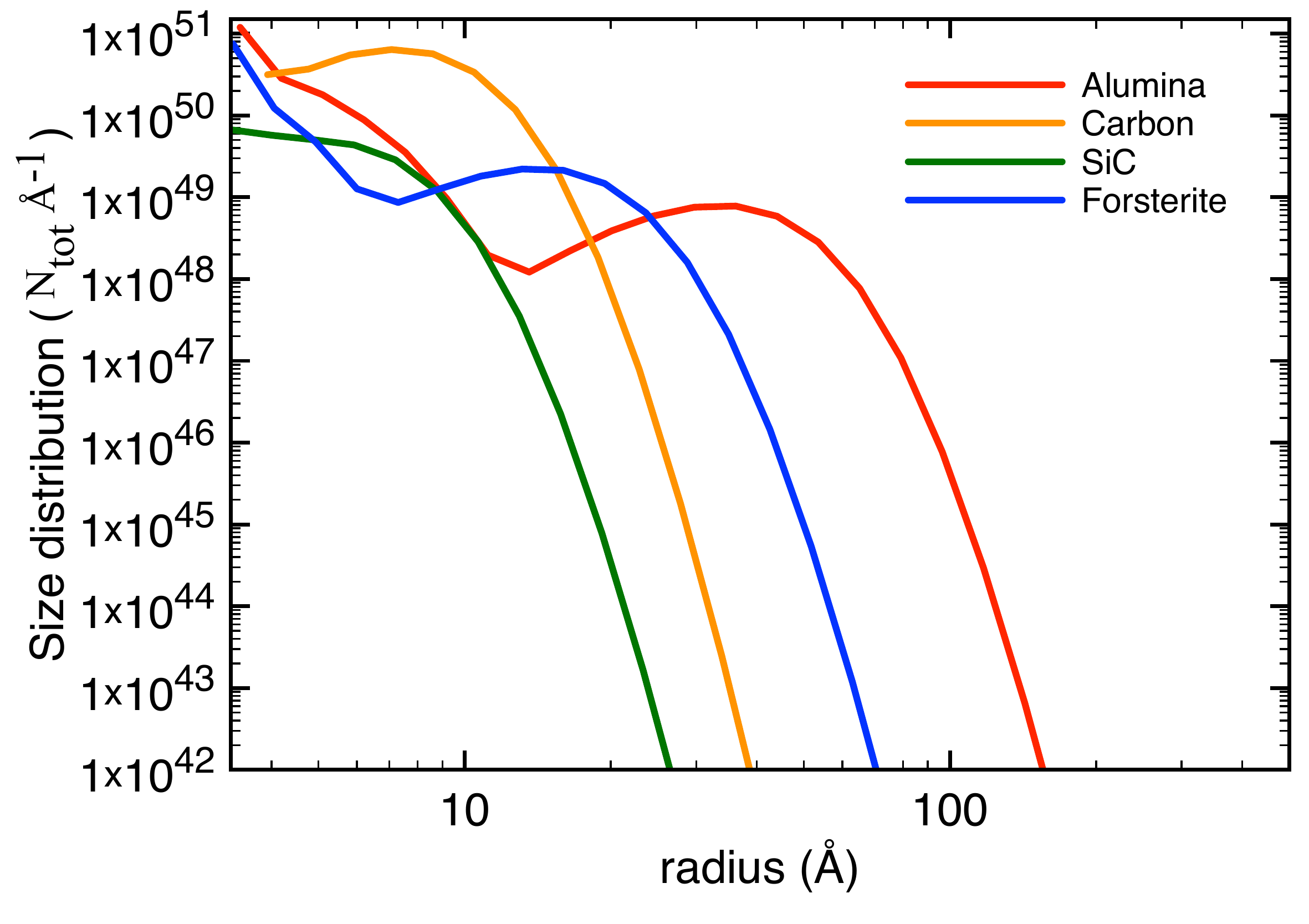}
\includegraphics[width=\columnwidth]{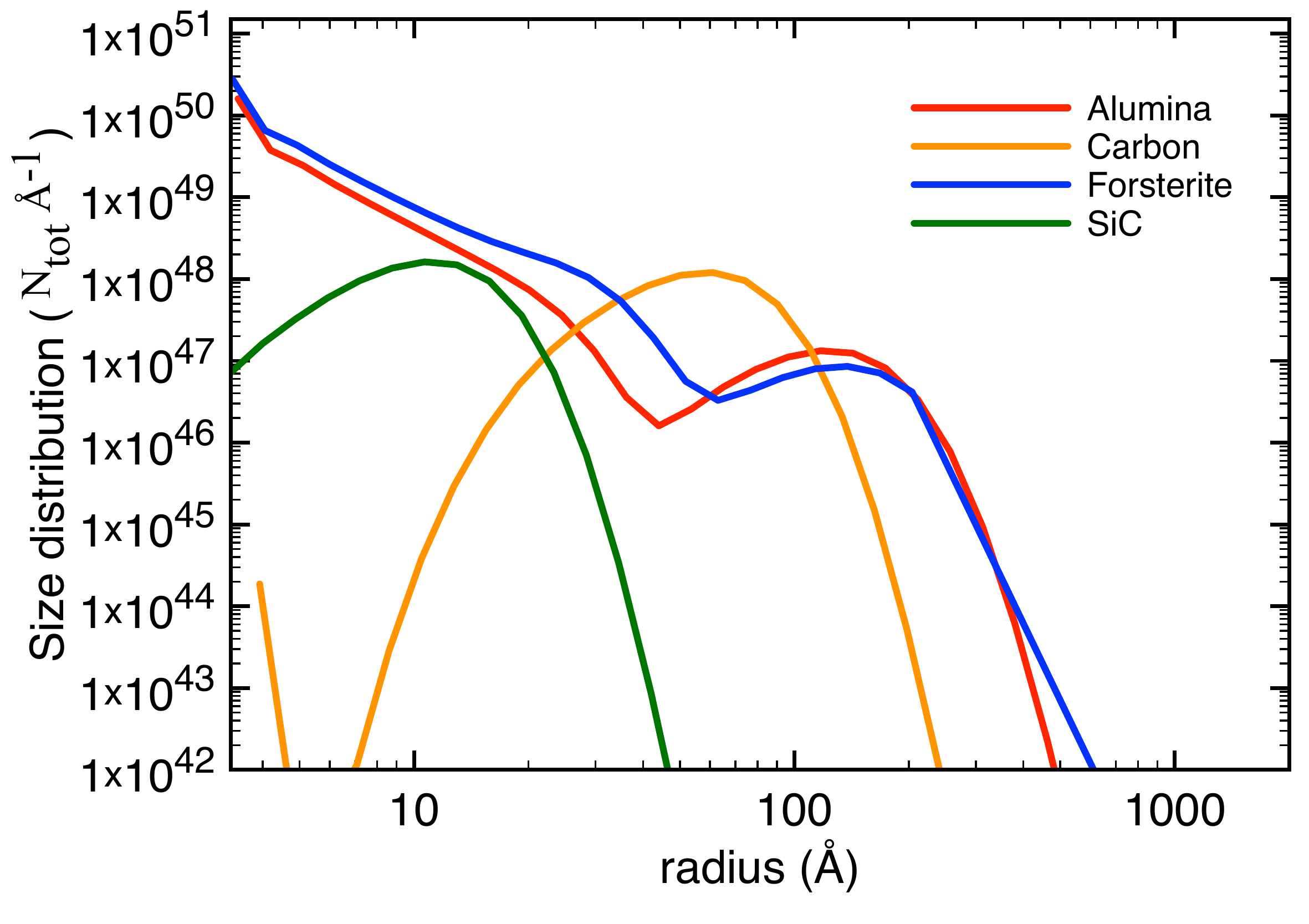}
\caption{Grain size distributions for the prevalent types of dust that form in SN ejecta at day 4 000 post-explosion. Top: Type~II-b SN ejecta appropriate to the Cas~A progenitor (case x200); Bottom: Type~II-P SN ejecta (case x2 000). Both SN progenitors have a mass of 19 \Ms.}
\label{fig0}
\end{figure}

\begin{table}
\centering      
\caption{Radius $a$ of the largest dust clusters formed from the gas phase and total dust mass summed over all ejecta zones in the SN at day 4 000 post-explosion.}    
\label{tab1}                     
\begin{tabular}{l  c c  c c}         
\hline\hline                  
Dust type & $a$ (\AA) & \multicolumn{2}{c}{Mass (\Ms)} \\
\hline
& & x200 & x2 000 \\

Forsterite &3.33 &3.1 $\times$ 10$^{-3}$ &1.1 $\times$ 10$^{-2}$\\
Alumina &3.45 &1.7 $\times$ 10$^{-2}$&1.4 $\times$ 10$^{-2}$\\
Carbon & 3.93&7.0 $\times$ 10$^{-3}$&7.1 $\times$ 10$^{-3}$\\
Silicon carbide &2.70&3.1 $\times$ 10$^{-4}$&9.1 $\times$ 10$^{-5}$ \\
\hline  
Total mass& & 2.7 $\times$ 10$^{-2}$  & 3.2 $\times$ 10$^{-2}$ \\
\hline                                    
\end{tabular}
\end{table}
We consider four dust types in the present study: silicate, alumina, carbon, and silicon carbide. Small dust clusters nucleate out of the gas phase to form dimers with forsterite and enstatite stoichiometry (Mg$_4$Si$_2$O$_8$ and Mg$_2$Si$_2$O$_6$, respectively), dimers of alumina (Al$_4$O$_6$), the smallest fullerene cage detected in the laboratory, C$_{28}$ (\cite{dun12}), and dimers of SiC, (SiC)$_2$. These small molecular clusters are then considered as so-called seeds that further coalesce and coagulate to form large dust grains. We follow the gas-phase chemistry, which is coupled with the dust condensation during 4 000 days after explosion, and derive dust grain size distributions and masses for a gas number density enhancement of 200 (hereafter x200) and 2 000 (hereafter x2 000). The grain size distributions obtained for the dust types that we considered are shown in Figure \ref{fig0}, and the final dust masses are given in Table \ref{tab1}, along with the radius of the small dust molecular clusters from which condensation starts. For the x200 case, the grains are small, with distributions peaking at radii in the range $20-70$~\AA. For the x2 000 case, grains are larger and the distributions peak at radii in the range $60-200$~\AA, except for SiC, which is present in the form of very small grains with radii less than $15$~\AA. In both cases, alumina and forsterite form the largest grains when compared to carbon and silicon carbide, because the oxygen-rich dust form at earlier post-explosion time and in denser ejecta zones than carbon-rich dust, which forms at late post-explosion times in the outermost zone of the ejecta (SC15).


\subsection{Clump model}
\label{CasA}

\begin{figure}
\centering
\includegraphics[width=\columnwidth]{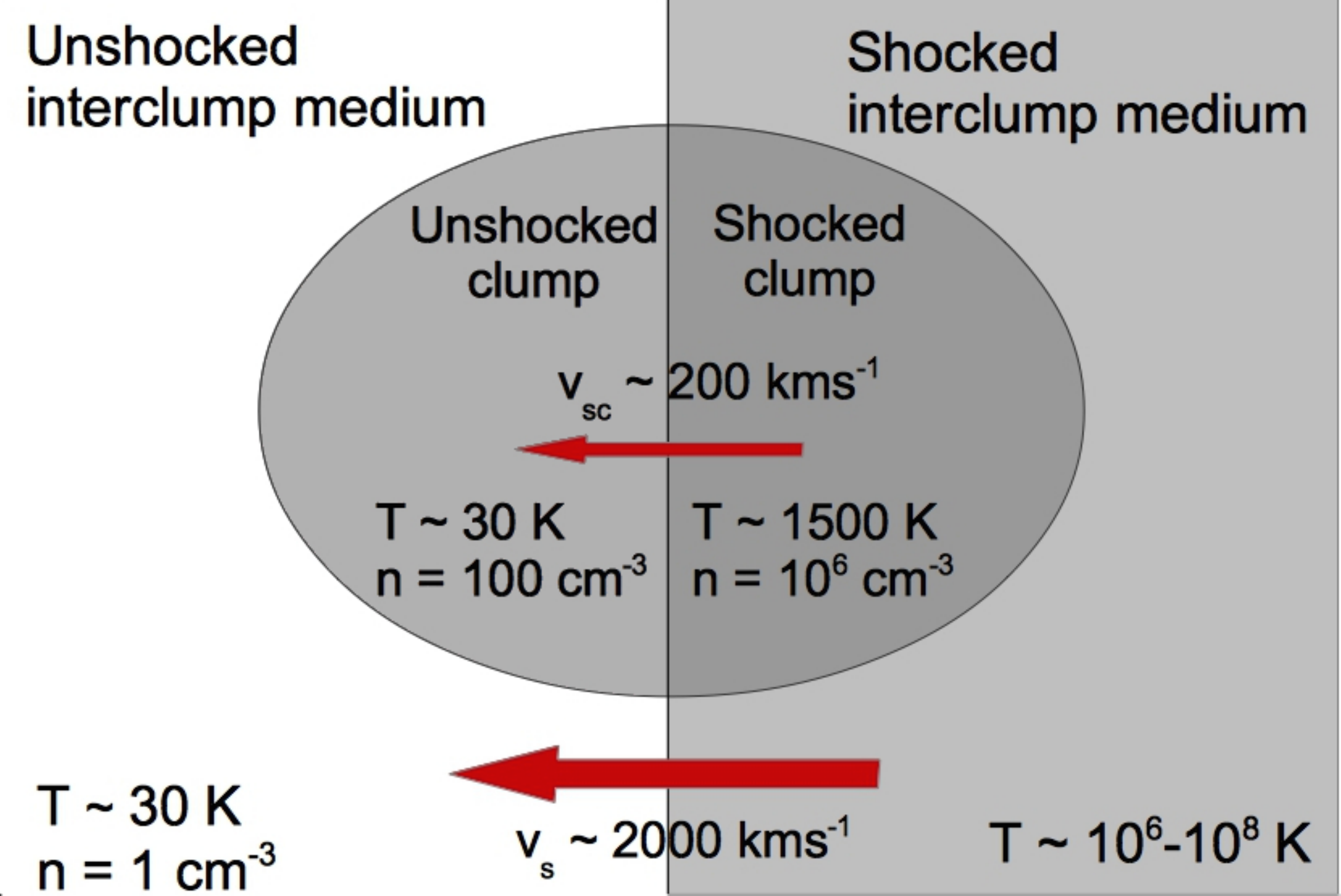}
\caption{Schematic representation of a shocked clump. The reverse shock is represented by the black line at the centre of the diagram and moves from right to left, as indicated by the red arrows. Pre-shock and post-shock values of the main gas parameters of the clump and inter-clump medium are indicated.}
\label{fig1}
\end{figure}

We describe the oxygen-rich core and the carbon-rich ejecta zone as ensembles of dense clumps imbedded in a diffuse inter-clump medium. The clumps and the associated inter-clump medium have a similar chemical composition. The dust content of clumps is described by initial grain size distributions as given in Figure \ref{fig0}. The clumps are crossed by the RS with reduced velocities of 100, 140, and 200 \kms\ within the clumps, and are subsequently disrupted. The gas conditions and shock velocities assumed in a clump that is crossed by the RS are described in BC14, and the post-shock parameters are listed in their Table~5. We provide a schematic view of one clump in Figure~\ref{fig1}, where pre- and post-shock conditions correspond to those derived from BC14 for similar RS velocities. We study dust sputtering in one clump embedded in its inter-clump medium, assume sputtering proceeds in a similar way in all the clumps that belong to one zone, and extend the results to the entire oxygen-rich core and the carbon-rich zone to derive the total mass of surviving dust. 

According to Klein et al. (1994), the cloud-crushing time, which is the characteristic time over which the clump is crushed by the RS, is given by 

\begin{equation}
\label{eq0}
\tau_{cc} = {a_c \over {V_s}} =  {\chi^{1/2} a_c \over {V_{sc}}}, 
\end{equation}

where $a_c$ is the radius of the clump, $V_s$ and $V_{sc}$ are the shock velocity in the inter-clump medium and within the clump, respectively (as shown in Figure \ref{fig1}), and $\chi$ is the density contrast between the clump and the inter-clump medium. Hydrodynamical simulations of shocked clumps in SNRs show that a reverse shock with velocity in the range 1 000~$-$~5 000~\kms\ is able to disrupt clumps with $\chi$ = 100 $-$ 1 000 on a time scale of $\sim$ 3 $\times \tau_ {cc}$ (\cite{sil10}). The clump disruption then releases dust grains into the diffuse and hot inter-clump medium. The inter-clump medium temperature is still subject to debate. The analysis of CHANDRA X-ray data of Cas~A gives typical inter-clump gas temperature between 10$^7$ K and 3 $\times$ 10$^7$ K for the region where the reverse shock is active (\cite{hwang12}). On the other hand, theoretical models consider different inter-clump gas temperatures. For example, Silvia et al. (2010) derive an inter-clump gas temperature of $\sim 10^6$ K from their hydrodynamical description of clump destruction by fast shocks, while Nozawa et al. (2010) use a temperature range of $10^6-10^7$~K in their homogeneous, shocked ejecta gas in Cas~A. Finally, Micelotta \& Dwek (2013) use a temperature of 10$^8$ K in their description of thermal sputtering in Cas~A. 

We model the non-thermal sputtering of dust grains within the clump and the later thermal sputtering in the inter-clump medium once the clump has been disrupted by the shock. To assess the effect of the inter-clump gas temperature on the results, we use temperatures in the range $10^6 - 10^8$ K, but choose temperatures of 10$^7$ K and 3 $\times$ 10$^7$ K for the Cas~A study. We consider the two over-density factors x200 and x2 000 for the clumps compared with the homogenous SN ejecta case as previously outlined. Both factors result in a ratio of clump/inter-clump number densities $\chi$ in the range $20-200$, as derived in BC14. These $\chi$ values are consistent with those derived from the observations of atomic lines in the shocked oxygen-rich clumpy structures present in Cas A, for which pre-shock gas densities in clumps of $\sim 100$ \cmc\ are derived (\cite{doc10}). 

\section{Dust sputtering}
\label{sput}
There are several possible destructive processes that are active on dust grains: grain disruption, vaporisation, and sputtering (\cite{drai79, jo96}). Grain disruption is caused by shattering as a result of grain-grain collisions. This process reduces the size of a grain but does not return material from the grain to the dust phase. Grain destruction processes include vaporisation and sputtering. Vaporisation is the total destruction of a grain after a collision with another grain. Since the number density of dust grains is small in the pre- and post-shocked clump, the processes that involve grain-grain encounters are suppressed and vaporisation can be ignored. Sputtering is instead caused by the impact between a grain and the surrounding gas, and leads to the removal of atoms or molecules from the surface of the grain. The sputtering of dust grains in SNRs occurs through non-thermal sputtering and thermal sputtering. Thermal sputtering results from the impact of the grains with the hot gas, where the motion of the colliding species is controlled by the gas temperature. Non-thermal sputtering occurs when there exists a large relative velocity between the gas and the dust grains. In this study, we concentrate on the evolution of the grain size resulting from non-thermal and thermal sputtering, which prevail in the context of fast, non radiative shocks, such as the reverse shocks in SNRs.

Several studies have explored the sputtering of dust grains in shocks, either for interstellar environments (Shull 1977; Cowie 1978; Barlow 1978; Draine \& Salpeter 1979; McKee 1987; Tielens et al. 1994; Jones et 1994, 1996) or in SNRs (Nozawa et al. 2006, 2007, 2010; Bianchi \& Schneider 2007; Nath et al. 2008; Silvia et al. 2010, 2012). The sputtering yield corresponds to the average number of sputtered atoms in the target (i.e., dust grain) per incident collider (atom or ion), and is derived by assuming the collision cascade theory (\cite{sig81}). All studies use the formalism developed by Bohdansky (1984), who defines the backward sputtering yield at normal incidence $Y_i (E)$ that is caused by a projectile $i$ impacting a target atom in the grain with energy $E$ as
\begin{align}
\label{eq1}
Y_i (E) &= 4.2 \times 10^{14} \frac{S_i(E)}{U_0} \frac{\alpha _i (\mu _i)}{K \mu _i +1} \left[ 1 -  \left(\frac{E_{th}}{E}\right)^{2/3} \right] \times \nonumber \\
 &\qquad {}  \left(1 - \frac{E_{th}}{E} \right)^2,
\end{align}
where $Y_i(E)$ has the unit of released atoms per ion. In Equation~\ref{eq1}, $U_0$ is the surface binding energy in eV, and $\mu _i = M_d/M_i $, where $M_d$ and $M_i$ are the masses of the target atom and the incident particle, respectively. Finally, $E_{th}$ is the threshold energy that needs to be provided in the collision to induce sputtering, $S_i(E)$ is the nuclear-stopping cross-section in units of ergs cm$^2$, $\alpha_i(\mu_i)$ is an energy-independent function of $\mu_i$, and $K$ is a free parameter that is adjusted to reproduce experimental data  available on sputtering yields. 

According to Andersen (1981) and Bohdansky (1984), the threshold energies $E_{th}$ are approximately given by 
\begin{equation}
\label{eq2}
E_{th} = \begin{cases}
\frac{U_0}{g_i(1-g_i)} & \text{for } M_i/M_d \leq 0.3 \\
 8U_0\left(\frac{M_i}{M_d}\right)^{1/3} & \text{for } M_i/M_d > 0.3 \end{cases}
,\end{equation}
where $g_i = 4M_iM_d/(M_i+M_d)^2$ is the maximum fractional energy transfer in a head-on elastic collision. According to Sigmund (1981), the nuclear-stopping cross section $S_i(E)$ is given by 
\begin{equation}
\label{eq3}
S_i(E) = 4\pi a_{sc}Z_iZ_d e^2 \frac{M_i}{M_i + M_d} s_i(\epsilon _i),
\end{equation}
where $e$ is the elementary charge, and $Z_i$ and $Z_d$ the atomic number of the incident ion and the target atom, respectively. The screening length $a_{sc}$ for the interaction potential between the nuclei is expressed as
\begin{equation}
\label{eq4}
a_{sc} = 0.885 a_0 (Z_i^{2/3} + Z_d^{2/3}) ^{-1/2},
\end{equation}
where $a_0 = 0.529$~\AA\ is the Bohr radius. Following Matsunami et al. (1980), an approximation to the function $s_i(\epsilon _i)$ is given by
\begin{equation}
\label{eq5}
s_i(\epsilon _i) =\frac{3.441 \sqrt{\epsilon _i}ln(\epsilon _i + 2.718)}{1+6.35\sqrt{\epsilon _i} + \epsilon _i(6.882\sqrt{\epsilon _i} -1.708)}, 
\end{equation} 
where $\epsilon _i$ is the reduced energy expressed as
\begin{equation}
\label{eq6}
\epsilon _i = \frac{M_d}{M_i+M_d}\frac{a_{sc}}{Z_iZ_de^2}E.
\end{equation}
The value of $\alpha _i(\mu _i)$ depends on the approximation of the distribution of energy deposited in the target. From the first approximation made by Bohransky et al. (1984), Nozawa et al. (2006) derive the following functions for $\alpha _i(\mu _i)$ by comparing sputtering data for $0.3 \leq \mu_i \leq 56$

\begin{equation}
\label{eq7}
\alpha_i = \begin{cases}
 0.2 &\text{for } \mu _i \leq 0.5 \\
 0.1 \mu _i ^{-1} + 0.25(\mu _i-0.5)^2&\text{for } 0.5 < \mu _i \leq 1 \\  0.3(\mu_i - 0.6)^{2/3} &\text{for } 1   < \mu_i
 \end{cases}
.\end{equation}

\begin{figure}
\centering
\includegraphics[width=\columnwidth]{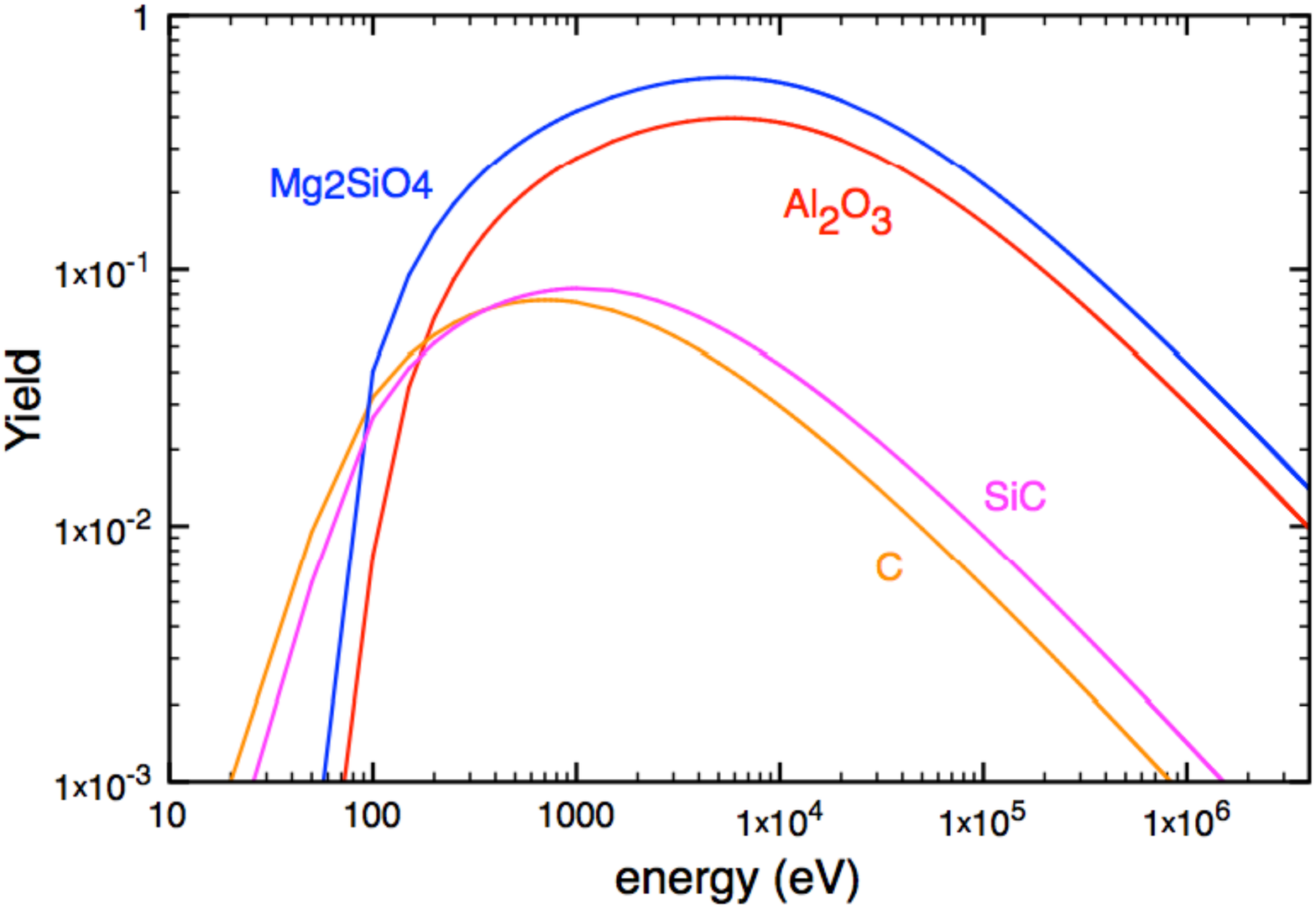}
\caption{Sputtering yield $Y_i(E)$ as a function of projectile energy $E$ for the dust species considered in this study. The yield for forsterite and alumina (carbon and silicon carbide) correspond to impact with O$^+$ (He$^+$).}
\label{fig2}
\end{figure}

The number $K$ is a free parameter that is determined by fitting the available experimental data of the sputtering yield for specific solids and by fitting simulated data for the grains for which no measured yields are available (\cite{tie94,noz06}). For alumina, carbon and silicon carbide, experimental data are available and used to determine $K$, whose values are listed in Table~\ref{tab2}, along with the other parameters entering the sputtering yield calculation. Forsterite has no sputtering yield data available in the literature for a large energy range. Following other studies (\cite{tie94,noz06}), we assume a fitting constant $K= 0.1$, which is that calculated for silica, SiO$_2$, while the binding energy of forsterite is an average adopted in studies of sputtering of interstellar silicates.

\begin{table}
\caption{The binding energy U$_0$, mean mass number of the atomic target M$_d$, mean atomic number of the atomic target Z$_d$ and the $K$ parameter, which enter the calculation of the yield $Y_i(E)$ given by Equation \ref{eq1} for the target materials considered in this study.}
\label{tab2}  
\centering
\begin{tabular}{c c c c c l}
\hline \hline  
Dust species & U$_0$(eV) & M$_d$& Z$_d$ & K & References\\
\hline
Mg$_2$SiO$_4$ &5.7 &20 &10 &0.1 & 8\\
Al$_2$O$_3$ &8.5 &20.4 &10 &0.08 & 3, 6, 7\\
SiC & 6.3& 20 & 10 & 0.3 & 1, 9, 3\\
C &4.0 &12 &6 &0.61 & 1, 2, 3, 4, 5\\
\hline
\end{tabular}
\tablefoot{
(1) Bohdansky 1978; (2) Roth et al. 1976; (3) Roth et al. 1979; (4) Rosenberg \& Wehner 1962; (5) Hechtl, et al. 1981; (6) Bach 1970; (7) Nenadovi{\' c} et al. 1990; (8) Tielens et al. 1994; (9) Behrisch et al. 1976.} 
\end{table}
 The sputtering yields for the dust material considered in this study are plotted in Figure \ref{fig2}. The sputtering ions were chosen according to the chemical composition of the zone where the dust grains form. Forsterite and alumina form in the oxygen-rich zones, where the most abundant ions are O$^+$, followed by Mg$^+$, while carbon and silicon carbide form in the outer zone, which is rich in He$^+$. While our sputtering yields for carbon and SiC agree well with those of Tielens et al. (1994), the yields for silicates and alumina are larger than values calculated by Tielens et al. (1994) and Nozawa et al. (2006) because we consider O$^+$ ions as the prevalent sputtering agent. 
\subsection{Non-thermal sputtering}

Following Draine \& Salpeter (1979) and Dwek et al. (1996), the erosion rate of grains is given by the variation of the grain radius $a$ with time, and calculated from the angle-averaged sputtering yield as follows:


\begin{align}
\label{eq8}
\frac{da}{dt} = &- \frac{m_{sp}}{2\rho _d} n \sum _i A_i \left(\frac{8kT}{\pi m_i}\right)^{1/2} \frac{e^{-s_i^2}}{2s_i}\nonumber \\
 & \times \int_{\epsilon _{th}}^{+\infty} \sqrt{\epsilon _i}e^{-\epsilon _i} sinh(2 s_i \sqrt{\epsilon _i}) Y_i(\epsilon _i) d\epsilon _i,
\end{align}

where $T$ is the gas temperature, $\epsilon _i = E_i / kT$, $\epsilon _{th} = E_{th}/ kT$ and $m_{sp}$ is the average mass of the sputtered atoms. 
\begin{figure}[h!]
\centering
\includegraphics[width=\columnwidth]{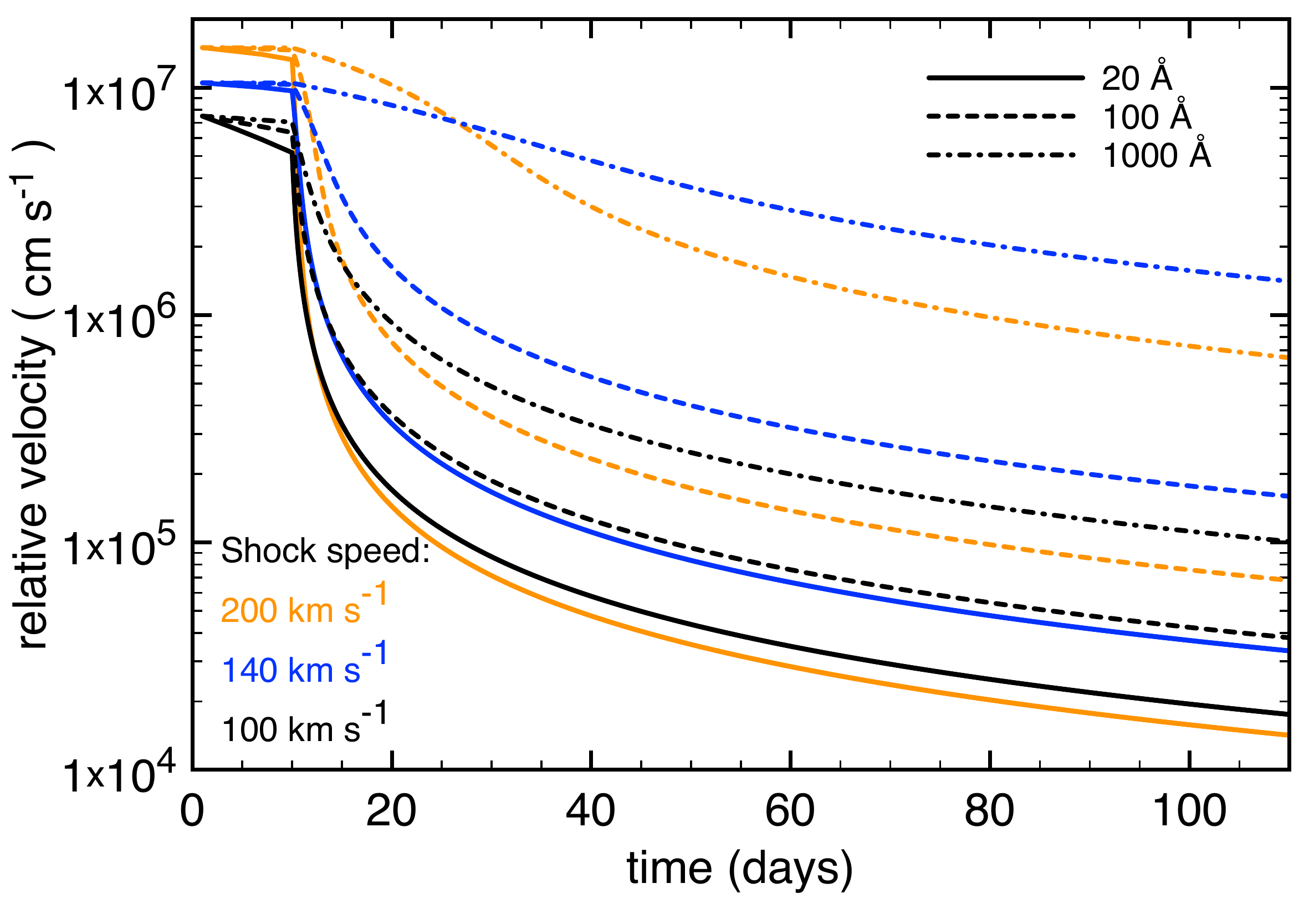}
\caption{Velocity of a silicate grain relative to the gas $v_d$ as a function of time in the post-shock gas for shock velocities $V_s =100, 140$ and $ 200$ \kms\ and grain radius $a=20, 100$ and  1 000~\AA.}
\label{fig3a}
\end{figure}

The quantity $s_i$ is expressed as

\begin{equation}
\label{eq11}
s_i^2 = \frac{m_iv_d^2}{2kT},
\end{equation}
where $m_i$ is the mass of gas species i, and $v_d$ is the dust velocity relative to the gas. In the limit where $s_i \rightarrow \infty$ (\cite{drai79}), which corresponds to the case where a hypersonic grain is eroded by non-thermal sputtering, the erosion rate is given by  

\begin{equation}
\label{eq9}
\frac{1}{n_{gas}} \frac{da}{dt} = - \frac{m_{sp} v_d}{2\rho_d} \sum_i A_i Y_i(E=0.5m_iv_d^2),
\end{equation}
where n$_{gas}$ is the gas number density and the other quantities have been previously defined. 
Because the dust crosses the shock front with a velocity $V_d$ equal to the shock velocity $V_s$, the dust velocity relative to the gas $v_d$ is initially equal to $v_d= V_d-v_{gas} = V_s - 1/4 V_s \simeq 3/4 V_{s}$ in the post-shock gas for an adiabatic shock. The velocity $v_d$ decreases with time as the dust grains are decelerated by colliding with the surrounding gas. If the gas is characterised by one temperature and the dust grain is spherical with radius $a$, the deceleration rate is given by 

\begin{equation}
\label{eq10}
\frac{dv_d}{dt} = - \frac{3n_{gas}kT}{2a\rho _d} \sum _i A_i G_i(s_i),
\end{equation}
where $\rho_d$ is the bulk density of the grain, and $A_i$ the abundance of gas species $i$ (\cite{drai79}). The function $G_i(s_i)$ is approximated as 
\begin{equation}
\label{eq12}
G_i(s_i)\approx \frac{8s_i}{3\sqrt{\pi}}\left( 1+ \frac{9\pi}{64}s_i^2\right)^{1/2}.
\end{equation}
According to Draine \& Salpeter (1979), this approximation is accurate within 1\% for $0<s_i< \infty$. The variation of $v_d$ with time according to Equation \ref{eq10} is shown in Figure \ref{fig3a} for several shock velocities within the clump and grain sizes for silicate grains. The drag forces that are due to collisions of the grain with the surrounding gas tend to bring the grain to rest and vary as $1 / (\rho_d \times a)$ (\cite{jo94}). Therefore, for a given material, small grains tend to decelerate faster than large grains, which are characterised by large $v_d$ values, as opposed to small grains. Consequently, the larger grains, which move hypersonically with respect to the gas, experience stronger NT sputtering in the post-shock gas than small grains, which are stationary within the clump.  
\begin{figure}[h!]
\centering
\includegraphics[width=\columnwidth]{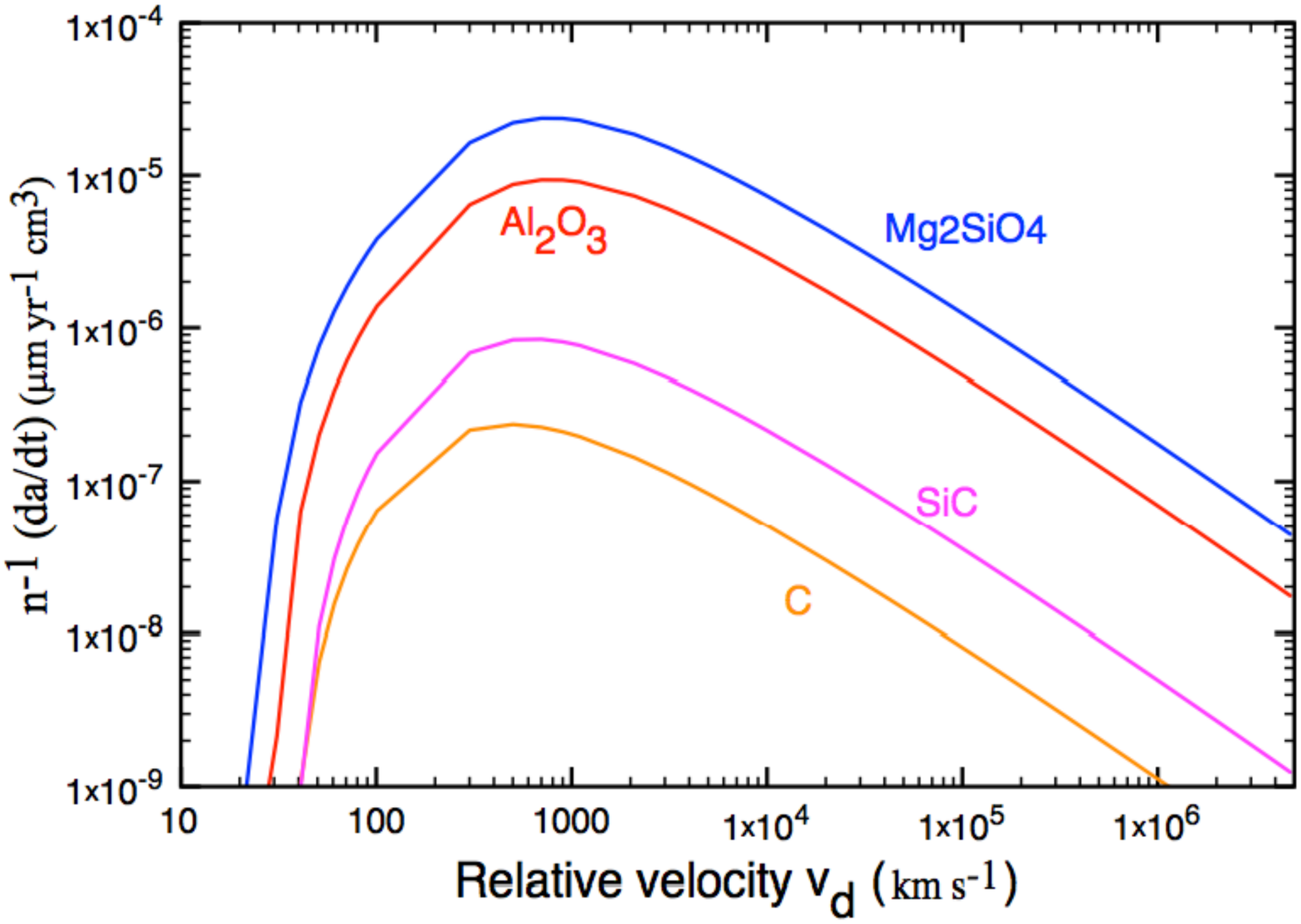}
\caption{Erosion rate for non-thermal sputtering as a function of dust velocities $v_d$ and types.}
\label{fig3}
\end{figure}

The low $v_d$ value for the 200 \kms\ shock shown in Figure~\ref{fig3a} results from the post-shock gas conditions of the model, which are given in Table~5 of BC14. The post-shock gas parameters for $V_{sc}=200$ \kms\ are derived from a different model than those used for the 100 \kms\ and 140 \kms\ shocks. For these two shock velocities, the $v_d$ variation reflects the fact that the post-shock gas velocity for the 140 \kms\ shock being smaller than that for the 100 \kms\ shock, the velocity of the grain relative to the gas $v_d(140)$ is larger than $v_d(100)$. However, the post-shock gas conditions for the 200 \kms\ shock are coming from a model which gives a post-shock gas velocity of 20 \kms, and thus a low $v_d(200)$ value.   

The erosion rates for the dust types of interest are calculated from Eq. \ref{eq9} as a function of the relative velocity $v_d$ and are shown in Figure \ref{fig3}. Each dust type is eroded according to the chemical composition of the zone from which it originates, with the most abundant element(s) taken as the main sputtering gas species. Carbon and SiC form in the external layer, and are thus eroded by He$^+$, while forsterite and alumina, which form in the oxygen-rich zone, are eroded by O$^+$ and Mg$^+$.\\

\subsection{Thermal sputtering}
If $s_i \rightarrow 0$, dust grains are stationary with respect to the gas and are eroded by colliding with the hot, surrounding plasma. The erosion rate for thermal sputtering is thus given by Equation \ref{eq8} when $s_i \rightarrow 0$
\begin{equation}
\label{eq13}
\frac{1}{n_{gas}} \frac{da}{dt}  = - \frac{m_{sp}}{2\rho_d} \sum_i A_i \left(\frac{8kT}{\pi m_i}\right)^{1/2} \int_{\epsilon _{th}}^{+\infty} \epsilon_i e^{-\epsilon_i} Y_i(\epsilon_i)d\epsilon_i, 
\end{equation}
where all quantities have been previously defined. 
\begin{figure}[h!]
\centering
\includegraphics[width=\columnwidth]{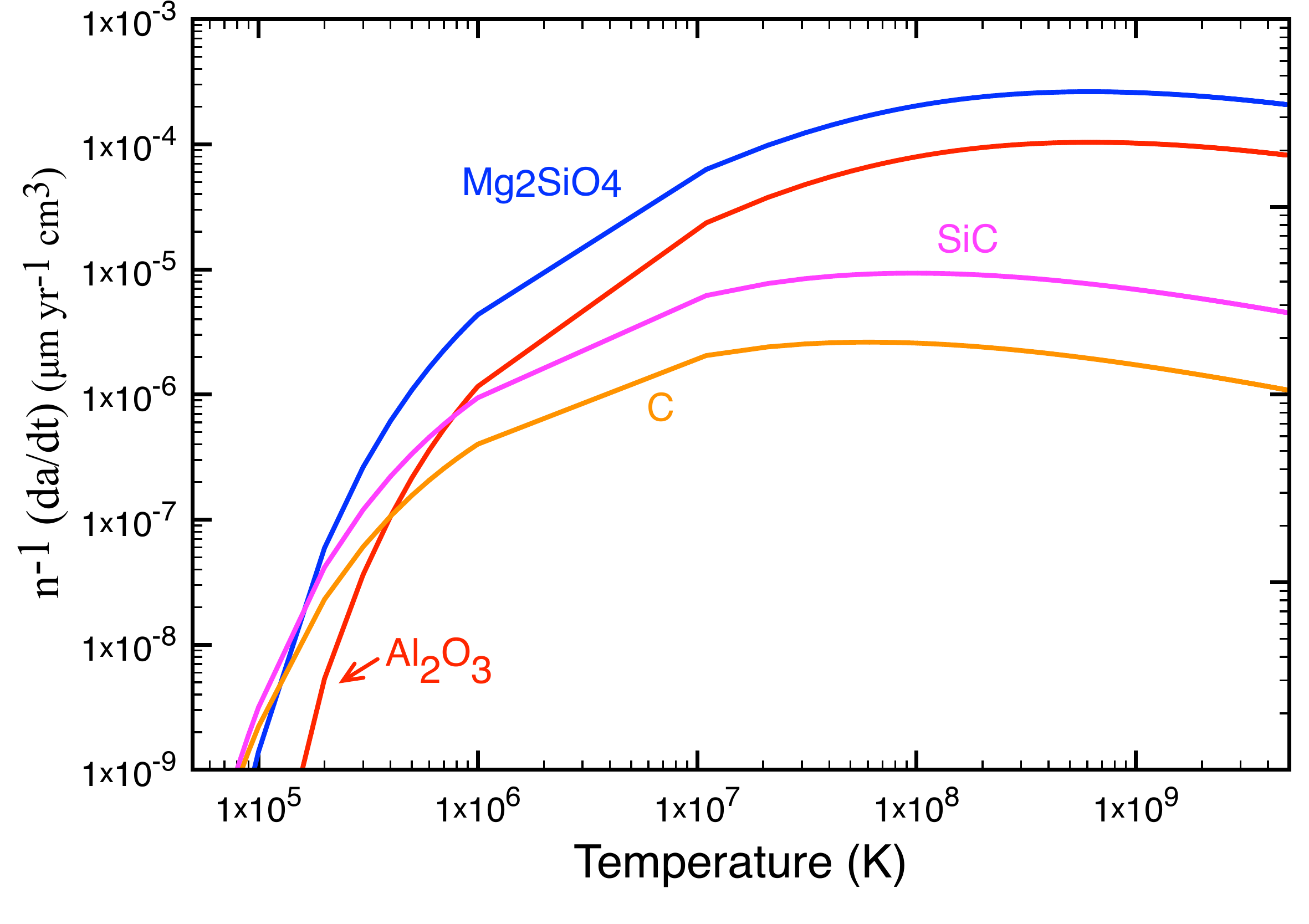}
\caption{Erosion rate for thermal sputtering as a function of gas temperature $T_{gas}$ and dust types.}
\label{fig4}
\end{figure}
Figure \ref{fig4} shows the erosion rates resulting from thermal sputtering for the dust types studied as a function of the gas temperature $T$. As for non-thermal sputtering, the composition of the zone where the dust forms determines which ions are prevalent for thermal sputtering, as we assume the clump expands in an inter-clump medium of similar chemical composition, as part of the SN ejecta. We see that the erosion rate take large values for gas temperatures in the range $10^6-10^8$ K, which are typical of SNRs.

The thermal sputtering in the inter-clump medium is considered until sputtering ceases. From Equation \ref{eq13}, we see the total thermal sputtering rate is proportional to the inter-clump gas number density $n_{gas}$. The expansion of the SN ejecta is homologous and is given by 
 
\begin{equation}
\label{dens}
n_{gas}(t) = n_{gas}(100) \times (t/100)^{-3}. 
\end{equation}
where n$_{gas}$(100) is the gas density at 100 days post-outburst. We derive the gas number density when thermal sputtering starts by assuming the time elapsed in units of years is

\begin{equation}
\label{time}
t_{thermal} = 340 + 3\times \tau_{cc},
\end{equation}
where 340 years is the age of Cas A and the second term is the clump-crushing time, where we choose a typical value of 100 years for $\tau_{cc}$ (\cite{sil10}). Thermal sputtering then starts at 640 years when the inter-clump gas number density given by Equation~\ref{dens} is $\sim 1$~\cmc, for an initial gas number density at day 100 post-explosion, as given by the x$200$ case of BC14. This value is assumed in our clump model and shown in Figure \ref{fig1}. The duration of thermal sputtering in the inter-clump gas was estimated to last $\sim 4 000$ years, when the inter-clump gas number density is $\sim 10^{-4}$~\cmc\ and the total sputtering rate becomes negligible. During that time period, we do not consider the cooling of the inter-clump gas, which is assumed to stay at a constant temperature in the range $10^6-10^8$~K.   

\begin{figure*}
 \centering
\includegraphics[width=6.07cm]{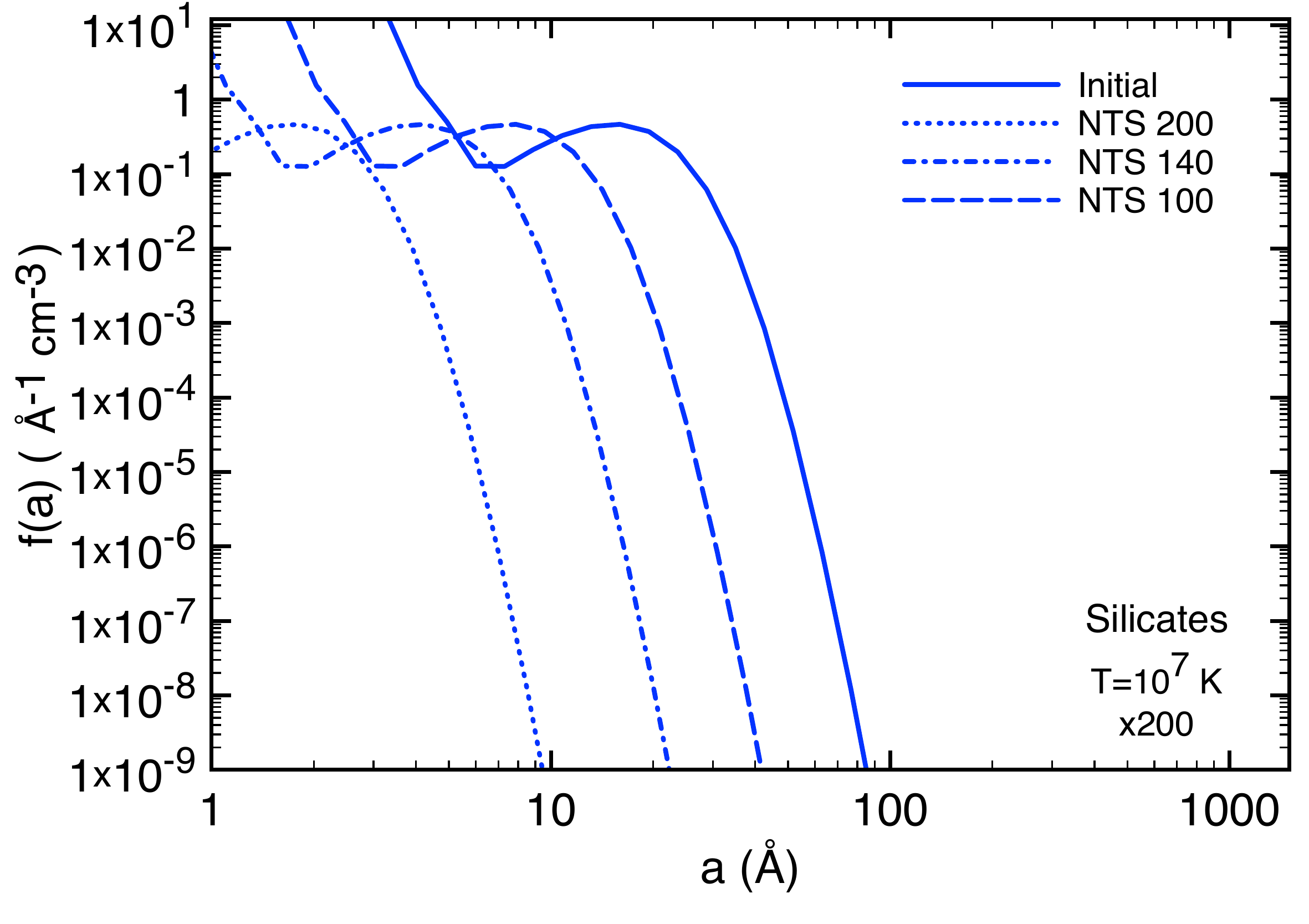}
\includegraphics[width=6.07cm]{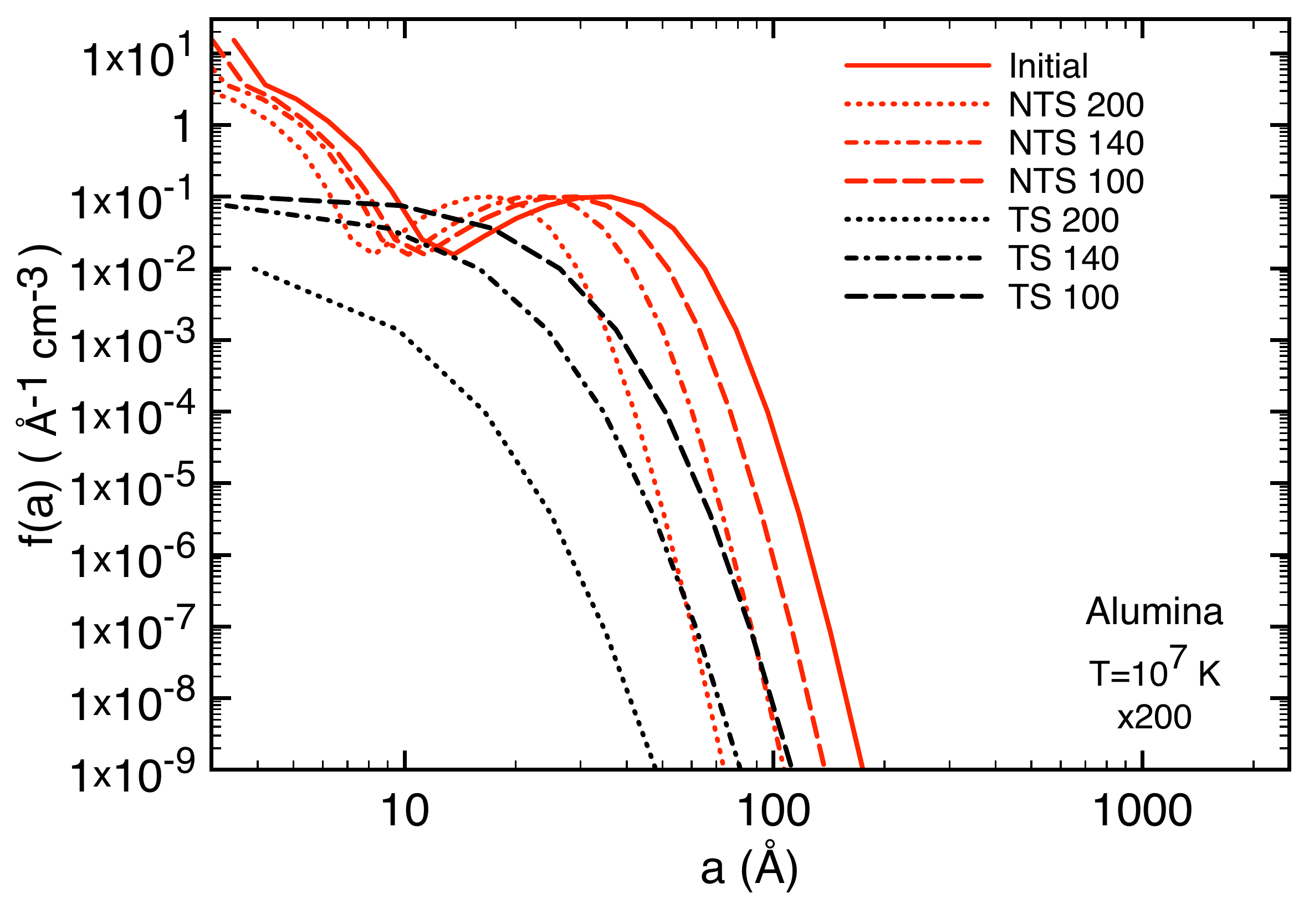}
\includegraphics[width=6.07cm]{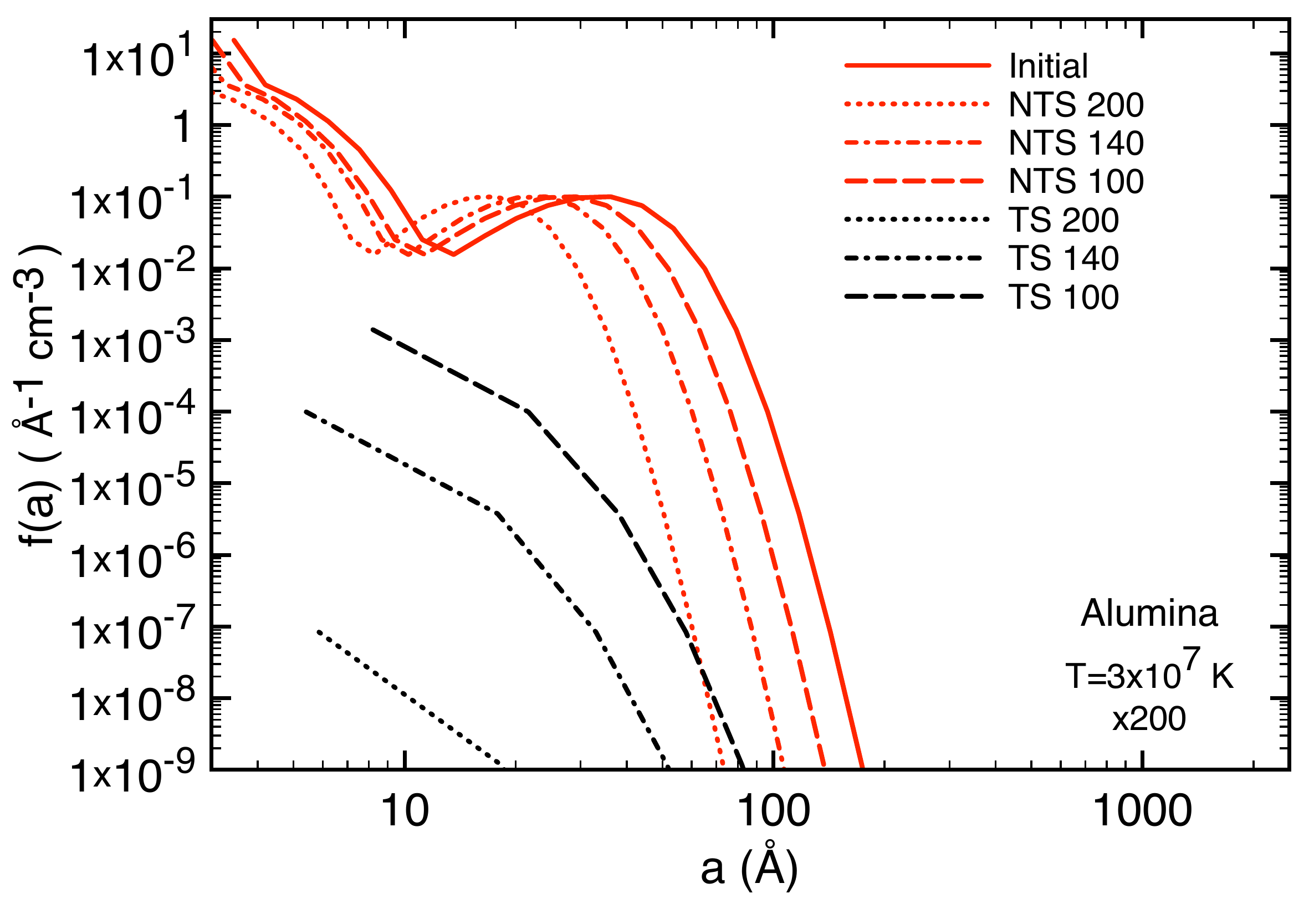}
\caption{Dust grains size distributions for the Cas A case (x200 clump) after non-thermal sputtering (NTS) in the oxygen-rich clump and thermal sputtering (TS) in the inter-clump medium. The shock velocities within the clump for all plots are 200 km s$^{-1}$ (dotted coloured lines), 140 km s$^{-1}$ (dotted-dashed coloured lines), 100 km s$^{-1}$ (dashed coloured lines). The initial size distribution is plotted with a full coloured line, while distributions after thermal sputtering are shown with black lines. Left: Silicates; Middle: Alumina with an inter-clump gas temperature of $10^7$~K; Right: Alumina with an inter-clump gas temperature of $3\times10^7$~K.} 
\label{fig5}
\end{figure*}

\section{Results}
\label{res}

\subsection{Sputtering in Cas~A}

\begin{figure}
\centering
\includegraphics[width=8cm]{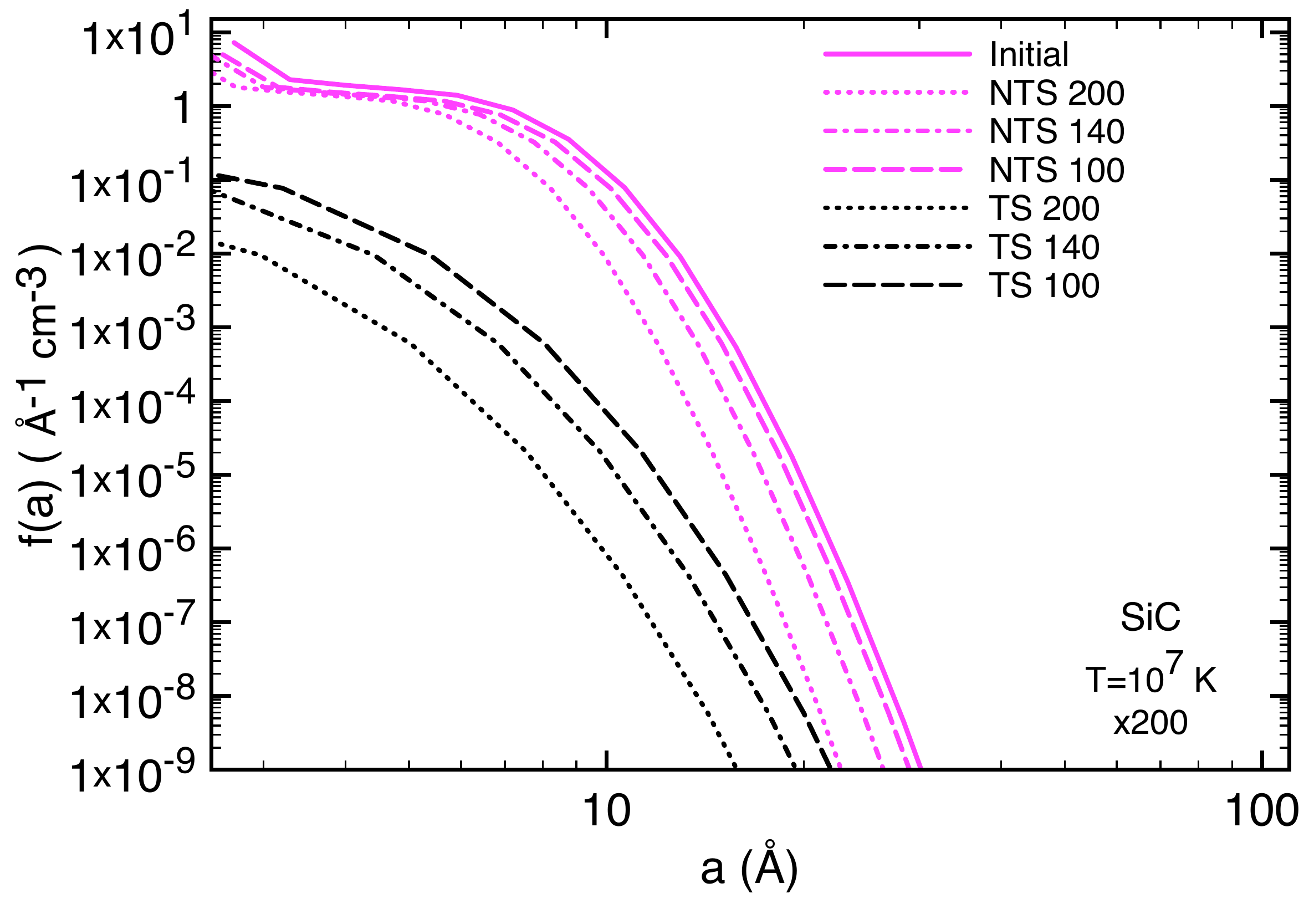}
\includegraphics[width=8cm]{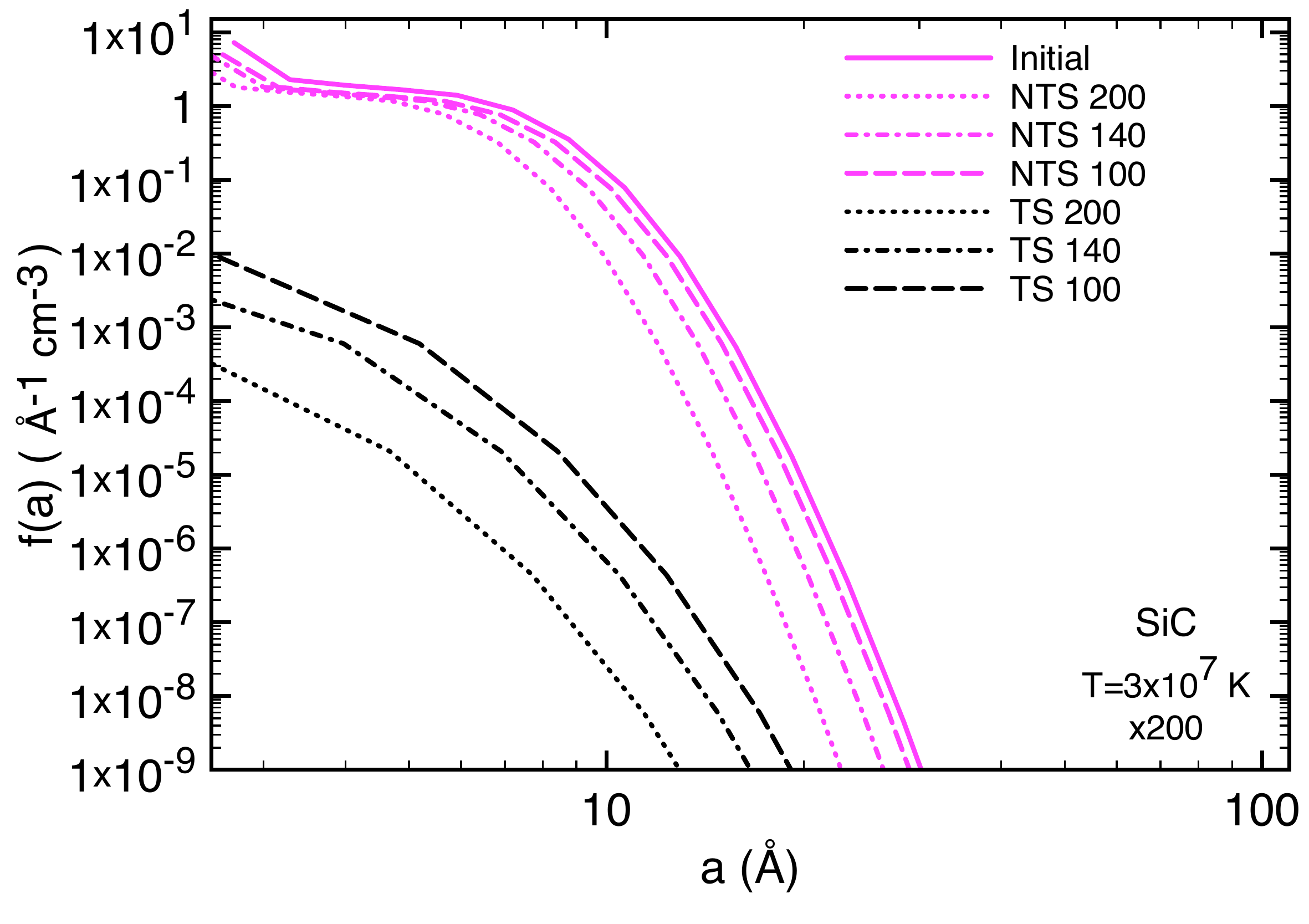}
\includegraphics[width=8cm]{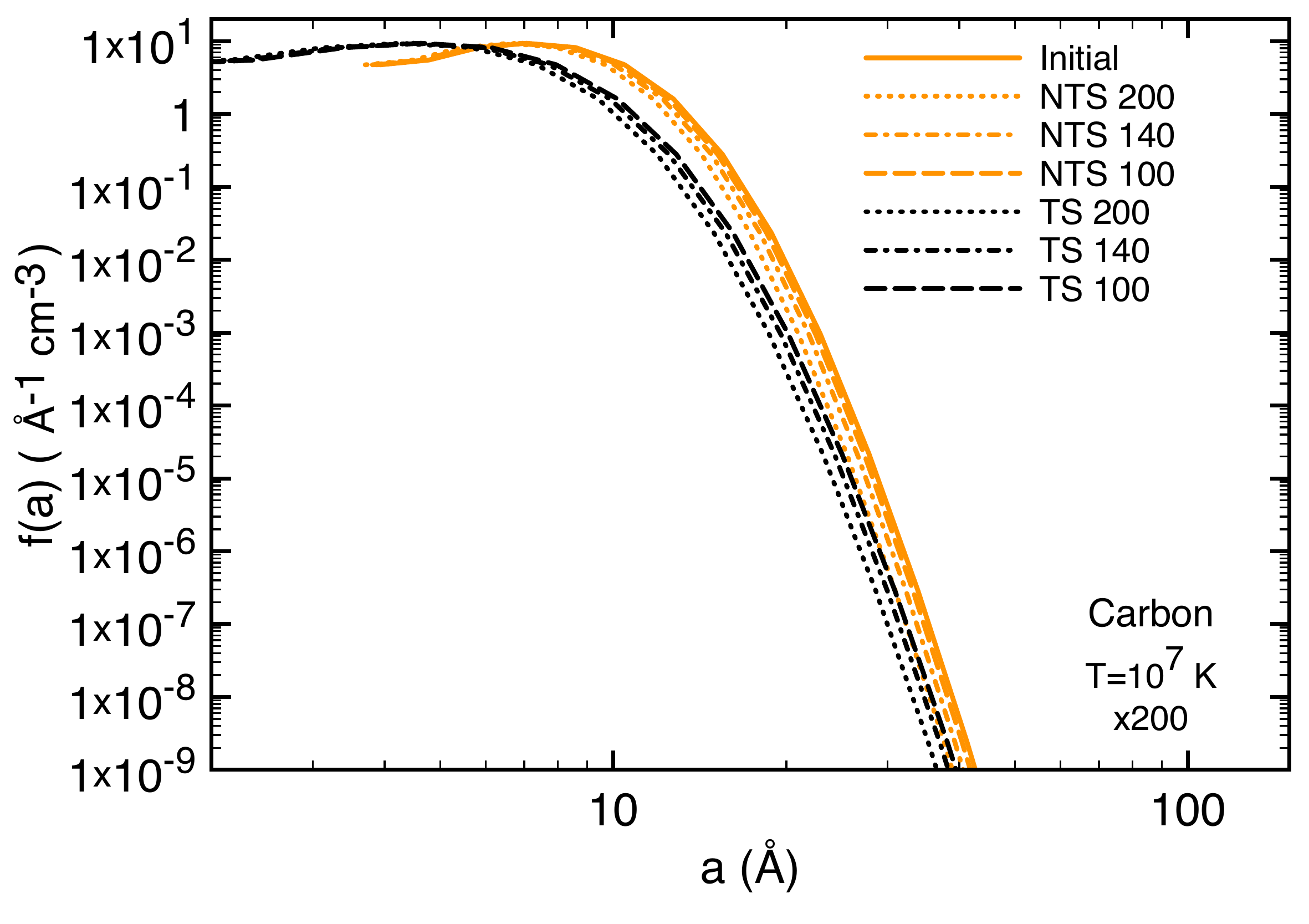}
\includegraphics[width=8cm]{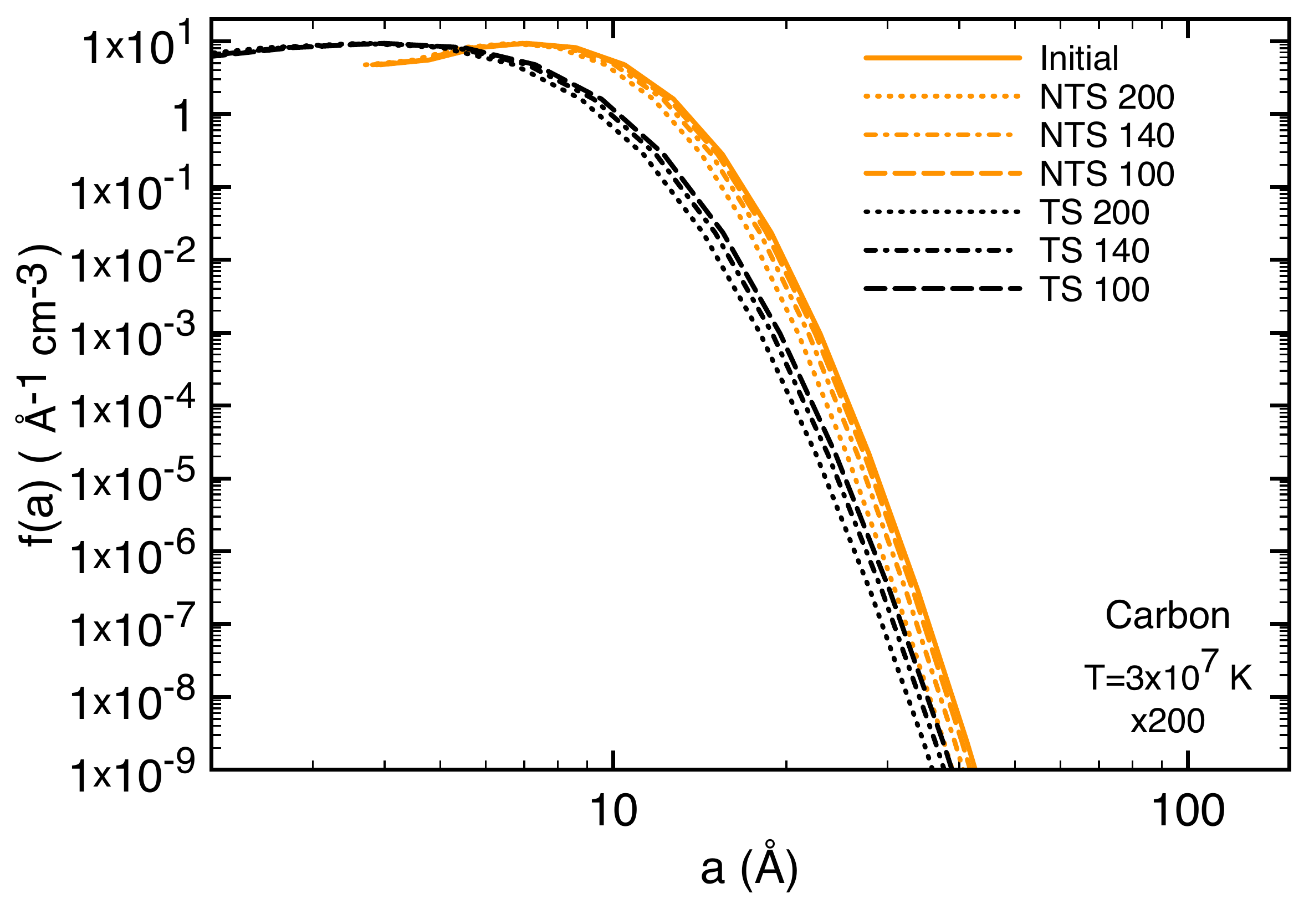}
\caption{Same as Figure \ref{fig5} for SiC and carbon dust sputtering in a carbon-rich clump and the inter-clump gas for Cas~A (x200 case). }
\label{fig6}
\end{figure}

As mentioned in Sect. \ref{CasA}, we consider non-thermal sputtering in clumps of various chemical compositions until the clump is disrupted over a few cloud-crushing times and dust thermal sputtering occurs in the hot inter-clump medium. Results for non-thermal and thermal sputtering are shown in Figures \ref{fig5} and \ref{fig6} for the four dust compositions of interest, the three RS velocities studied and the two inter-clump temperatures that characterise Cas A ($10^7$ K and $3 \times 10^7$ K). The total dust mass summed over all dust types and survival efficiency are summarised in Table \ref{tab3}. 

\subsubsection{Oxygen-rich clump}

The initial dust composition of the clump is that of the oxygen-rich core of the SN progenitor, where the prevalent dust components are alumina and silicates, and the non-thermal (NT) sputtering agents are O$^+$ and Mg$^+$ ions. For silicates, the initial grain distribution is severely skewed towards very small grains owing to NT sputtering in the clump for all RS velocities. The initial peak at $\sim 20$~\AA\ is shifted to $a$ values in the range $2-8$~\AA\ within the clump. Such very small grains (or large molecular clusters) are readily and totally destroyed by thermal sputtering in the inter-clump medium for all inter-clump gas temperatures once the clump is disrupted. This results in no final size distributions being shown in Figure \ref{fig5} for silicates. The situation is different for alumina for which the initial size distribution peaks at larger radius ($\sim 40$~\AA). Therefore, NT sputtering within the clump reduces the overall initial size distribution to smaller grains where the largest grains have radius of  $15-30$~\AA\ depending on RS velocities. Thermal sputtering in the inter-clump medium tends to flatten the size distribution because small grains are preferentially destroyed. For the smallest inter-clump temperature, thermal sputtering further reduces the grain sizes to leave a minor population of very small grains with $a$ in the range $2-20$~\AA. The destruction is more important for the larger inter-clump temperature, especially for the fast $200$~\kms\ shock, where almost all grains are destroyed. More generally, the gas and shock conditions that characterise Cas A are not conducive to the survival of a silicate grain population, while a small fraction of alumina grains may survive.  

\subsubsection{Carbon-rich clump}

We investigate sputtering in a knot that is characterised by the composition of the carbon rich zone of the Cas~A ejecta. We consider that both carbon and silicon carbide dust grains are the prevalent solids in the zone and study NT and thermal sputtering by He$^+$ ions. Results for Cas A are presented in Figure \ref{fig6}. Both SiC and carbon grains form in the ejecta with small sizes as shown in Figure \ref{fig0}, where the distributions peak at $\sim 7$~\AA\ and $9$~\AA, respectively. Such small grains suffer less NT sputtering in the clump than the larger alumina and silicate grains. For SiC grains, NT sputtering shifts the size distribution to the left with peaks in the range $5-7$~\AA. Thermal sputtering in the inter-clump gas further destroy the grains and produces a modest population of very small grains over time. Carbon suffers less NT and thermal sputtering than SiC because, as seen from Figure \ref{fig2}, carbon has a lower sputtering yield in the range $10^3-10^4$~eV, which results in lower erosion rates for NT and thermal sputtering. The initial grain size distribution is not much altered by NT sputtering and slightly shift to lower grain size as a result of thermal sputtering. A population of very small carbon grains that have a size in the range $4-8$ \AA\ will then survive the RS in the remnant phase. 

In general, sputtering is more severe in the clump and the inter-clump medium for the dust formed in the oxygen-rich core because the initial sputtering yields are higher owing to the more massive sputtering agents (see Figure \ref{fig2}). This results in more severe destruction of the oxygen-rich dust, which includes silicates and alumina grains. However, silicates and oxides are formed at early post-explosion time and in regions of higher gas density in the SN ejecta (Sarangi \& Cherchneff 2013, SC15), with the largest grains being of alumina. Therefore, in the case of alumina, the effect of sputtering by heavy ions is compensated for by the larger initial grains, which allows some alumina grains to survive the RS passage and thermal sputtering in the remnant. But overall, the results indicate that the amount of dust formed in Type~IIb~SNe is severely destroyed in the remnant. For Cas~A, the surviving dust mass after 4 000 years of sputtering in the remnant phase ranges between 6~\% and 11~\% of the initial mass. 

\begin{table}
\centering      
\caption{Total dust mass (in \Ms) and survival efficiency left after NT and thermal sputtering as a function of RS velocities ($V_{sc}$) and inter-clump temperature (T) for the Cas~A remnant (case x200, interclump T =$10^7-3\times 10^7$~K).  
The total initial dust mass before sputtering is $2.7\times~10^{-2}$~\Ms. }   
\label{tab3}                      
\begin{tabular}{c   c c c c}         
\hline\hline     
\noalign{\vskip 0.7mm}                    
 T(K) & M(NT) & $\epsilon_{NT} $ & M(Thermal) & $\epsilon$  \\
\hline
\multicolumn{5}{c}{ $V_{sc}=100$ \kms}  \\
\hline
\noalign{\vskip 0.7mm}   
 $10^7$ &  $1.6 \times 10^{-2}$ &58.3\% &$2.9 \times 10^{-3}$ & 10.8\% \\$3\times10^7$ &   &&$2.9 \times 10^{-3}$ & 7.4\% \\ 
 \hline
\multicolumn{5}{c}{ $V_{sc}=140$ \kms} \\
\hline
\noalign{\vskip 0.7mm}     
 $10^7$ &  $1.1\times 10^{-2}$ &41.0\% &$2.4 \times 10^{-3}$ & 8.6\% \\
 $3\times 10^7$ &  &&$1.6 \times 10^{-3}$ & 5.9\% \\
\hline
\multicolumn{5}{c}{ $V_{sc}=200$ \kms} \\
\hline
\noalign{\vskip 0.7mm}    
  $10^7$  &  $7.7\times 10^{-3}$ &27.9\% &$2.1 \times 10^{-3}$ & 7.7\% \\  $3\times10^7$  &  &&$1.5 \times 10^{-3}$ & 5.6\% \\
\hline                                    
\end{tabular}
\end{table}

\subsubsection{Current sputtering in Cas~A}

\begin{figure}

\centering
\includegraphics[width=\columnwidth]{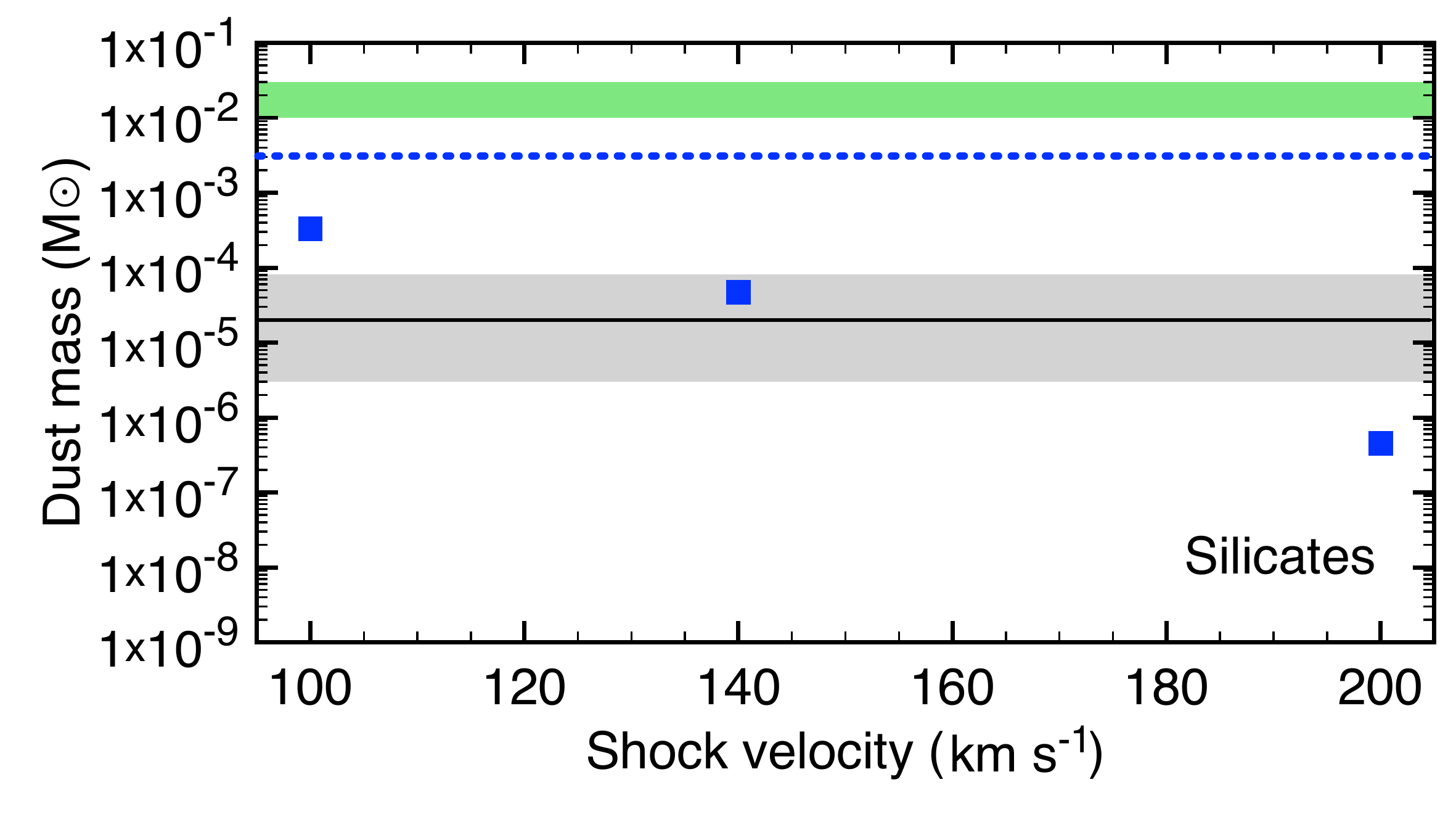}
\includegraphics[width=\columnwidth]{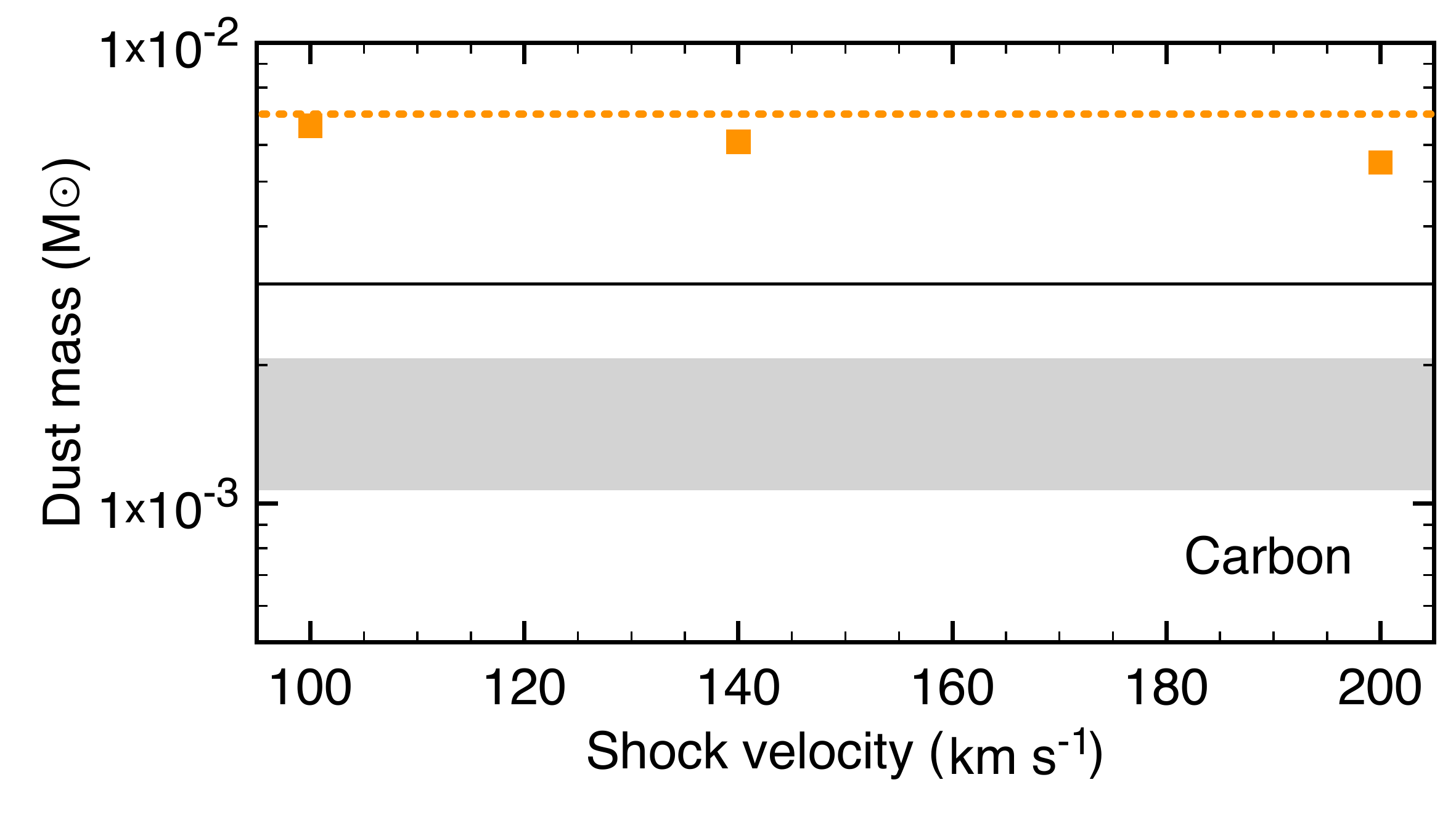}
\includegraphics[width=\columnwidth]{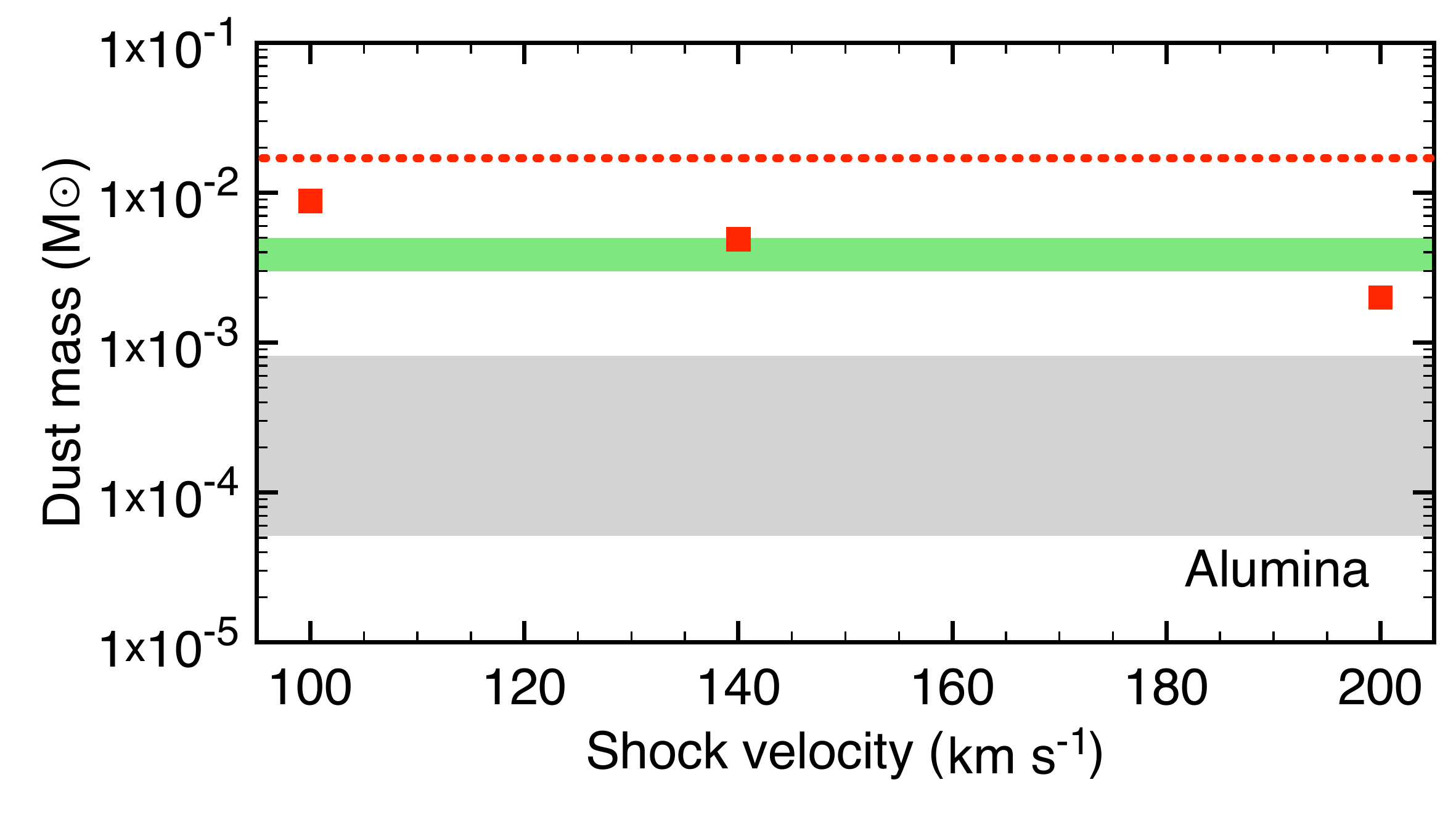}
\caption{Modelled masses of carbon, alumina, and silicate dust (coloured square) at present in Cas~A compared with the masses derived from Spitzer data by Rho et al. (2008) (grey shaded area) and Arendt et al. (2014) (green shaded area). The initial dust mass formed by the supernova is given as a dotted coloured line. Dust masses derived for the SNR 1E 0102.2-7219 in the Small Magellanic Cloud from {\it Spitzer} data (\cite{sand09}) are also shown (full back line). }
\label{fig7}
\end{figure}

Observations of the IR continuum in Cas A provide information on the current dust content in the remnant. Mapping in the mid-IR with {\it ISO} and {\it Spitzer} provided information on the warm dust (\cite{ar99,dou01,rho08,ar14}), while colder dust was probed with Herschel PACS observations at 70 \mic, 100 \mic,\ and 160 \mic\ (\cite{bar10}). These observations indicate that a large fraction of the dust is in the form of oxygen-rich dust, namely silicates and alumina.  
Figure \ref{fig7} shows the dust mass derived for our x200 case for silicate, carbon, and alumina dust `at present', i.e., the dust masses processed by the non-thermal sputtering within the oxygen- and carbon-rich clumps crossed by the RS. We expect thermal sputtering to occur after a few cloud-crushing times $\sim 3 \times \tau_{cc}$ (see Sect. \ref{CasA}). This time is comparable to the Cas A remnant age and indicates that thermal sputtering should not currently proceed. The dust masses estimated by Rho et al. (2008) and Arendt et al. (2014) from Spitzer and Herschel data are shown for comparison, as well as the initial dust masses formed in the supernova clumps for our x200 case before NT sputtering, as listed in Table \ref{tab1}. We also show the dust masses derived for 1E 0102.2-7219, a remnant similar to Cas A in the Small Magellanic Cloud (\cite{sand09}). 

According to our model, the current situation on the dust mass evolution in Cas A can be depicted by any mass comprised between the initial mass values and the final masses at times when NT sputtering ends. All three RS velocities can be considered because they correspond to various level of clumpiness within the remnant. Our results for the oxygen-rich dust, alumina and silicate, and carbon dust agree rather well with the masses deduced by Rho et al. (2008) from {\it Spitzer} data. All modelled masses after NT sputtering are within a factor of ten or less, compared with values derived from observations for all RS velocities. 

Our results and those of Arendt et al. (2014) for dust grains that formed in the O-rich core agree well for alumina, while there is some discrepancy for silicate dust. In their study, Arendt et al. analyse {\it Spitzer} data according to the dominant emission lines in regions of the remnant. They find that silicates are the prevalent dust in regions dominated by Ar II and Ar III lines, while alumina dust best characterises the regions rich in Ne II emission lines. Because Cas~A has kept some memory of the initial chemical stratification of its SN progenitor (\cite{del10, ise12, mil13}), this result is not surprising. As shown in BS14, alumina is the primary dust component in the Ne-rich zone 2, with a mass that is about eight times that of silicates in this zone. On the other hand, silicates form at first in the only Ar-rich region, zone 1B, and is the primary dust component in this zone. Their total silicate mass derived from the Ar II and Ar III line regions is $\sim 4.5 $ times larger than the mass of silicate ejecta dust listed in Table~\ref{tab1}. However, their silicate mass is much larger than our NT-sputtered silicate mass. Their mass value is derived by using a specific silicate stoichiometry with a low Mg content (Mg$_{0.7}$SiO$_{2.7}$), a compound that is close to the pyroxene enstatite, (MgSiO$_3$). Although several other silicate types have been included in the fitting process, we note that the absorption coefficients of pyroxene silicates are smaller than those of olivine silicates in the $9-25$~\mic\ wavelength range (\cite{fab00}). This results in overestimating the derived dust mass as the mass is directly proportional to the inverse of the mass absorption coefficient of the various dust components. In chemical kinetic models of dust formation in SN ejecta, the mass fraction of enstatite is several orders of magnitude less than that of forsterite in the ejecta of Type~II-b and Type-II SNe, and pyroxene clusters are intermediates in the olivine cluster formation process (Goumans \& Bromley 2012; BC14; SC15). No explanation is provided by Arendt et al. as to why silicates close to a pyroxene stoichiometry should preferentially form in the SN that led to Cas~A. 

No information on carbon dust mass and location is provided by Arendt et al. (2014). The possible contribution of carbon dust in Ar II, III, and Ne II regions is discussed as the data on O-rich dust, specifically silicates, require the contribution of a featureless component in the fitting procedure. This contribution can also be provided by alumina grains according to the authors. A carbon contribution seems unlikely since carbon dust is essentially formed in the outermost mass zone of the ejecta and is not expected to be abundant in the Ar- and Ne-rich ejecta regions (SC15). 

Globally, our current sputtered masses for the various dust components that are derived for Cas A are in good agreement with those derived from {\it Spitzer} observations. This seems to indicate that dust in the clumpy ejecta of Cas A has, essentially, been processed  by NT sputtering within dense clumps. 

\subsection{Sputtering in Type~II-P supernovae}
\label{SNIIP}

\begin{figure*}
\centering
\includegraphics[width=6cm]{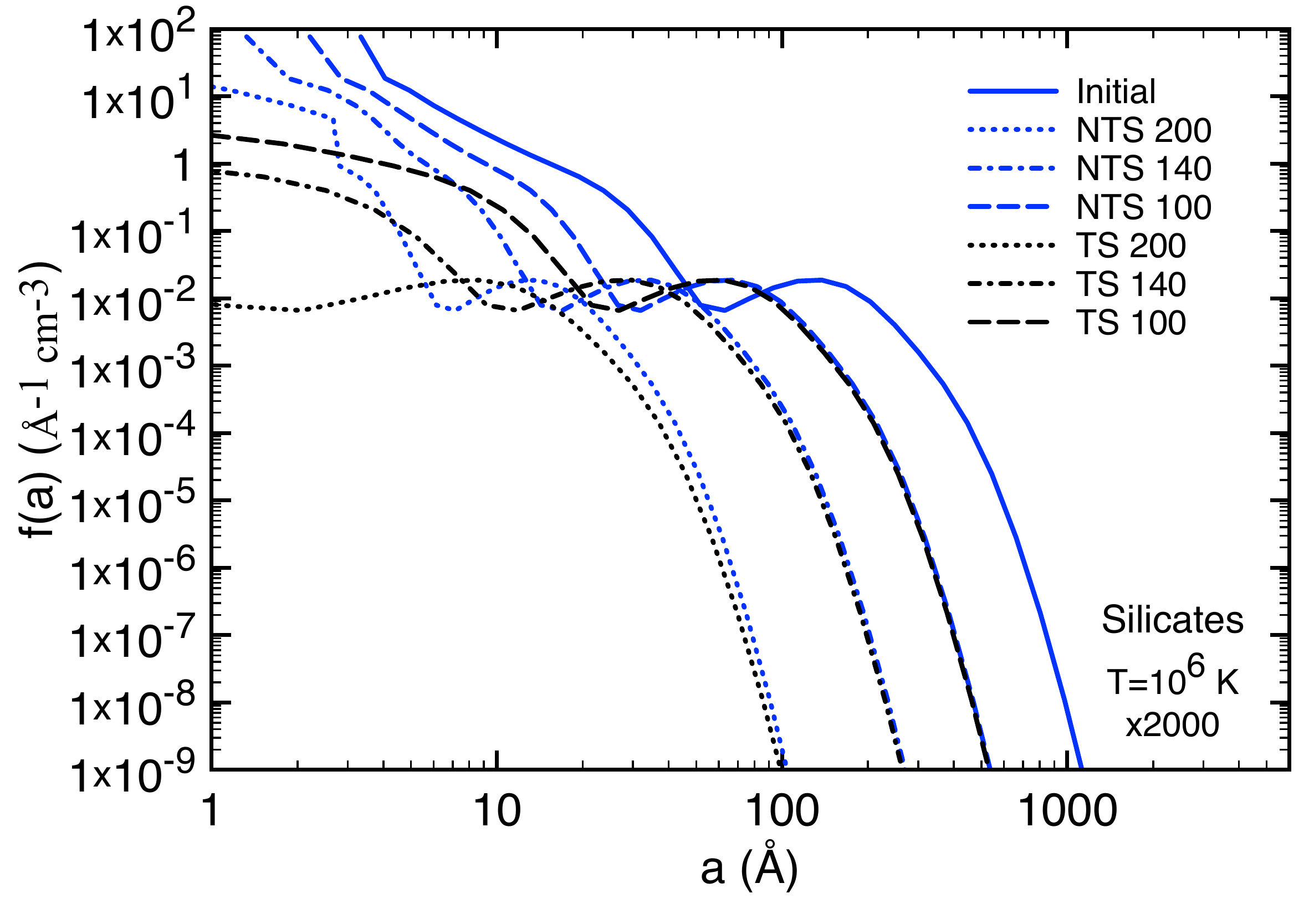}
\includegraphics[width=6cm]{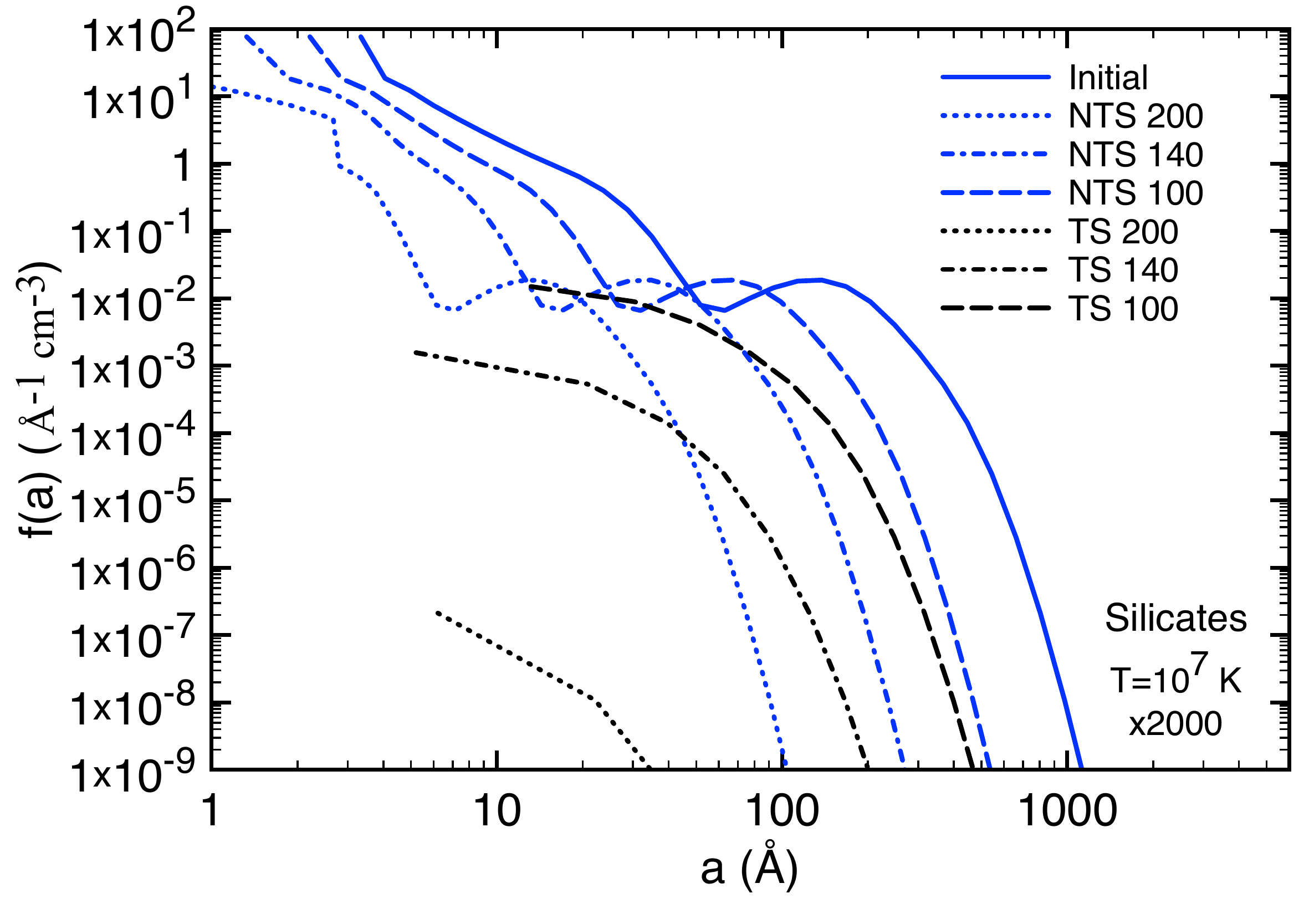}
\includegraphics[width=6cm]{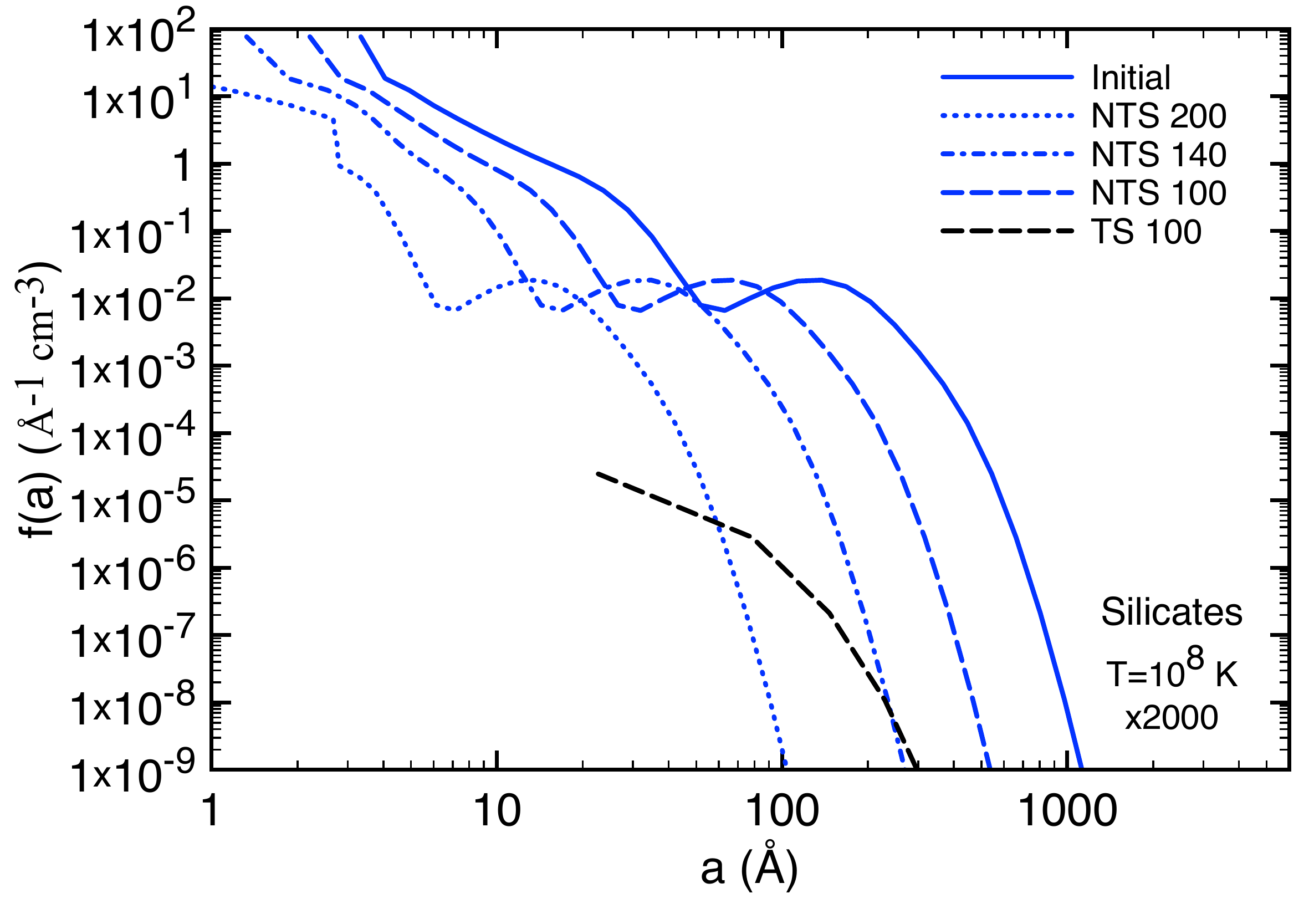}
\includegraphics[width=6cm]{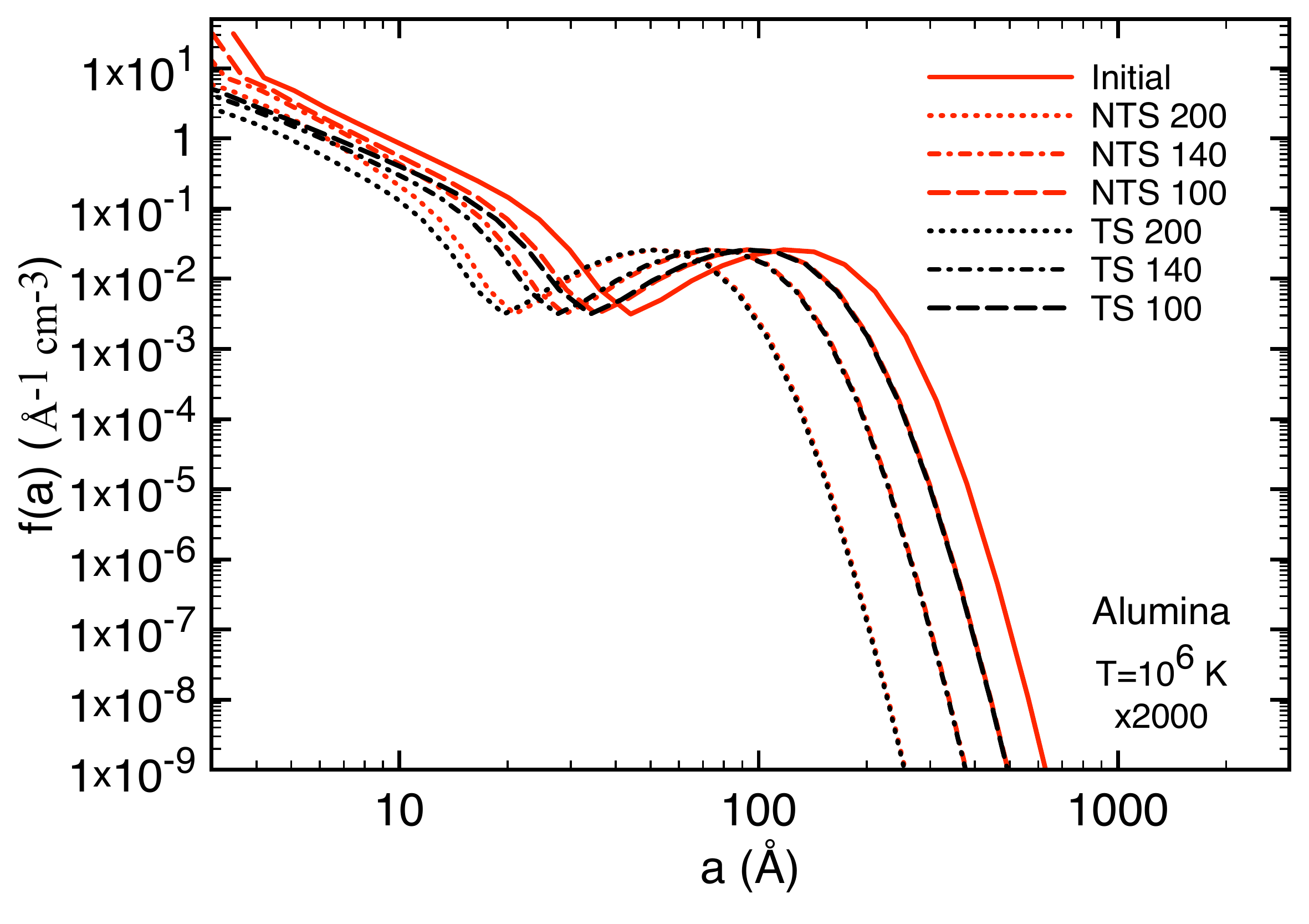}
\includegraphics[width=6cm]{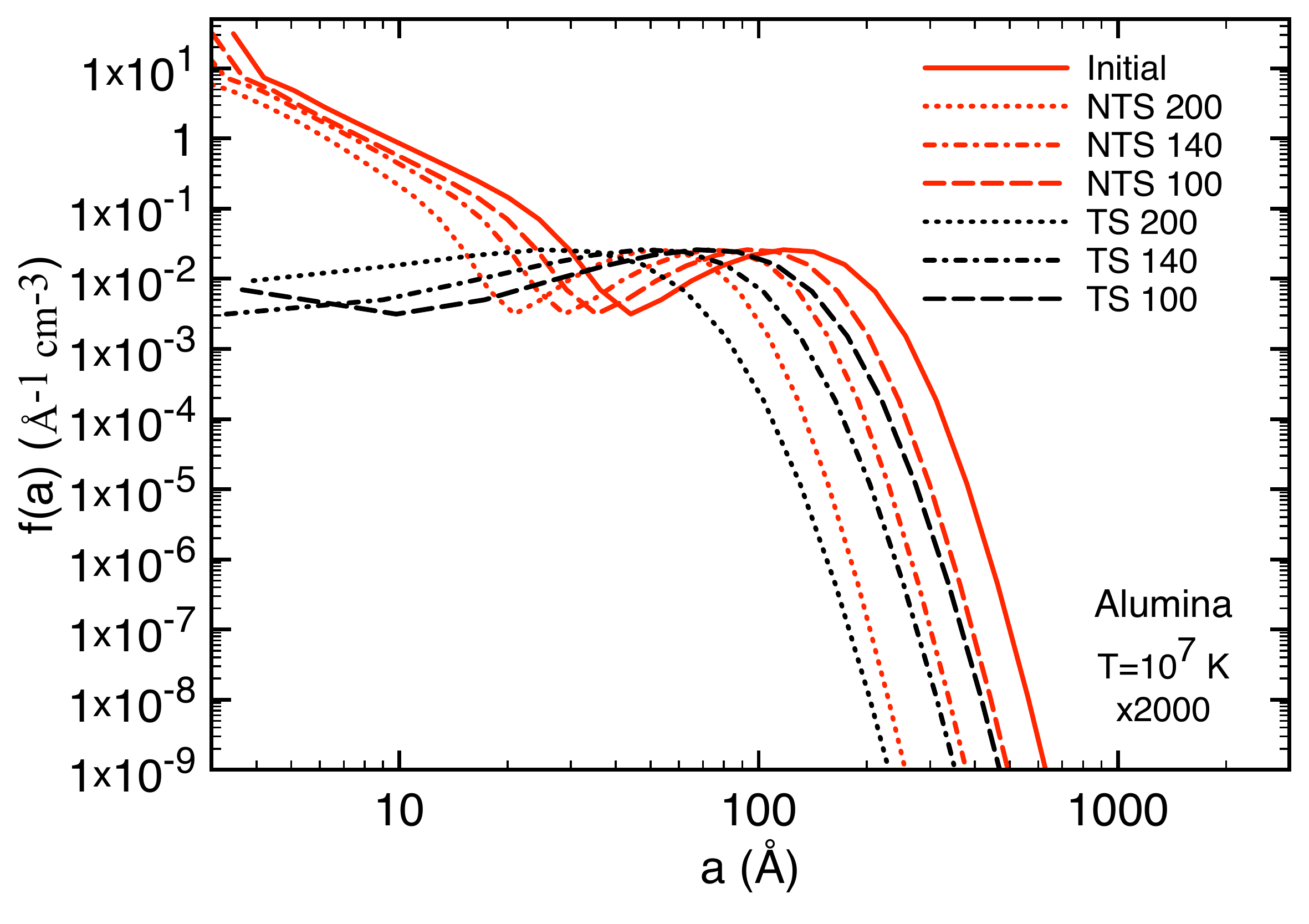}
\includegraphics[width=6cm]{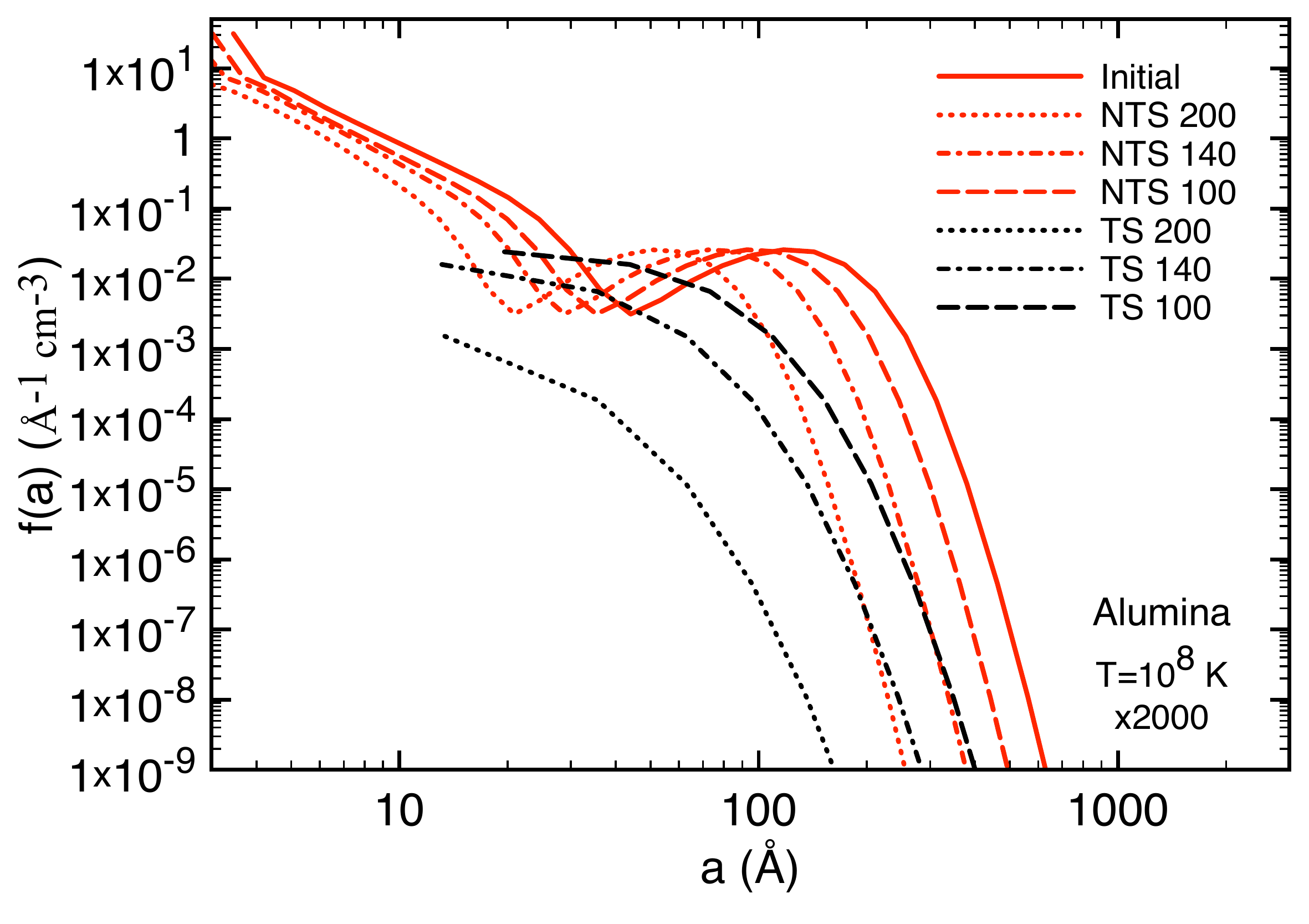}
\includegraphics[width=6cm]{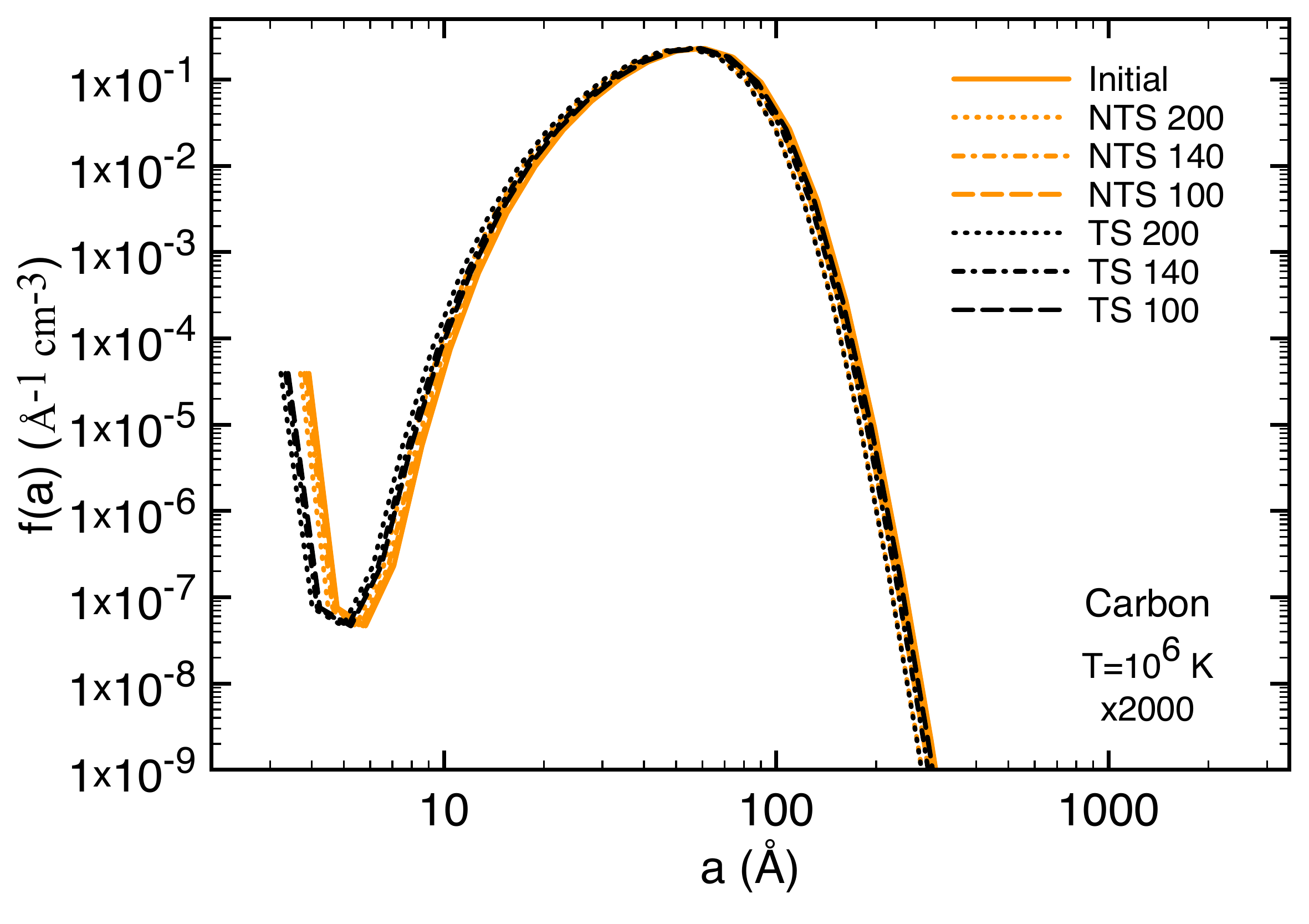}
\includegraphics[width=6cm]{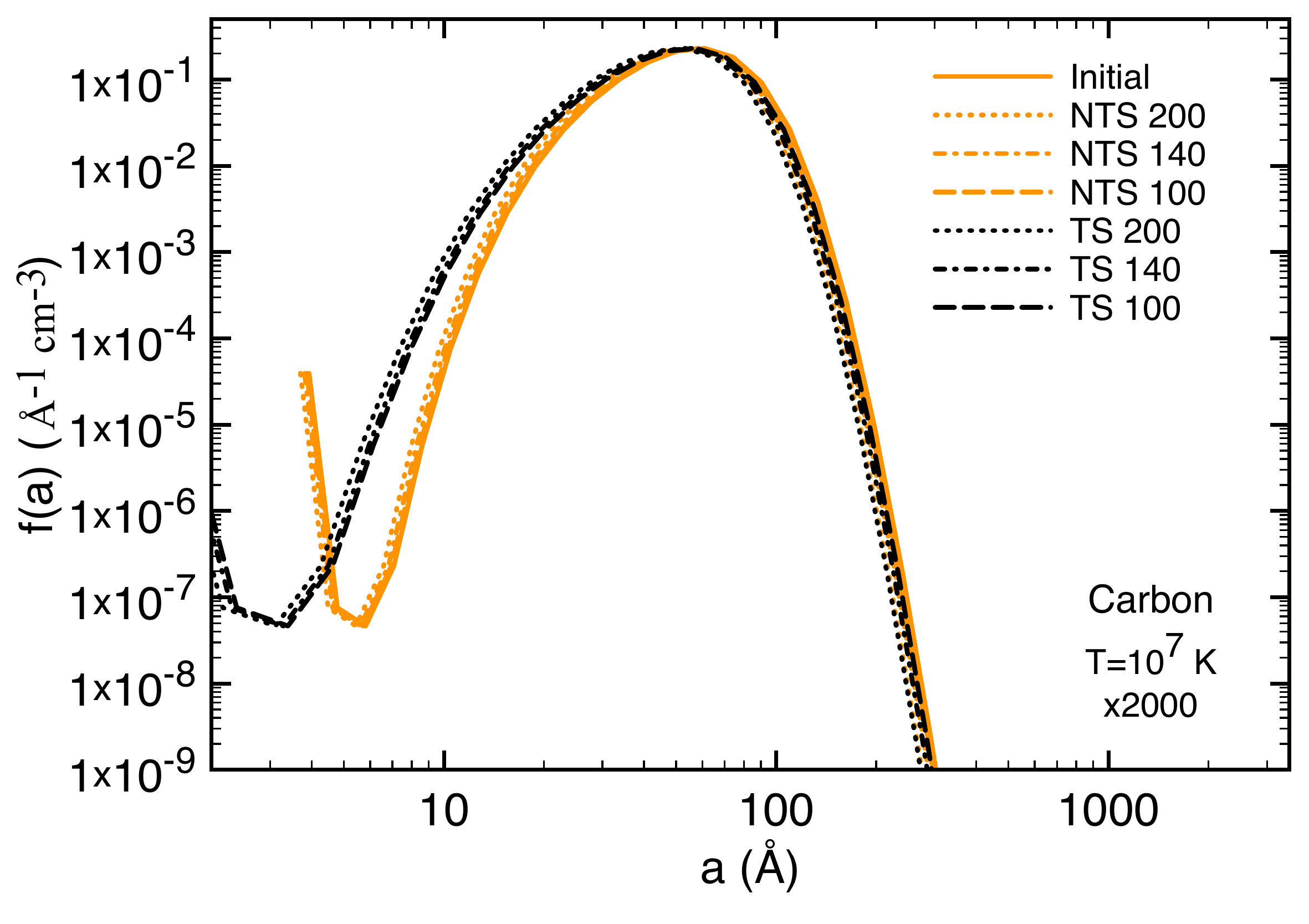}
\includegraphics[width=6cm]{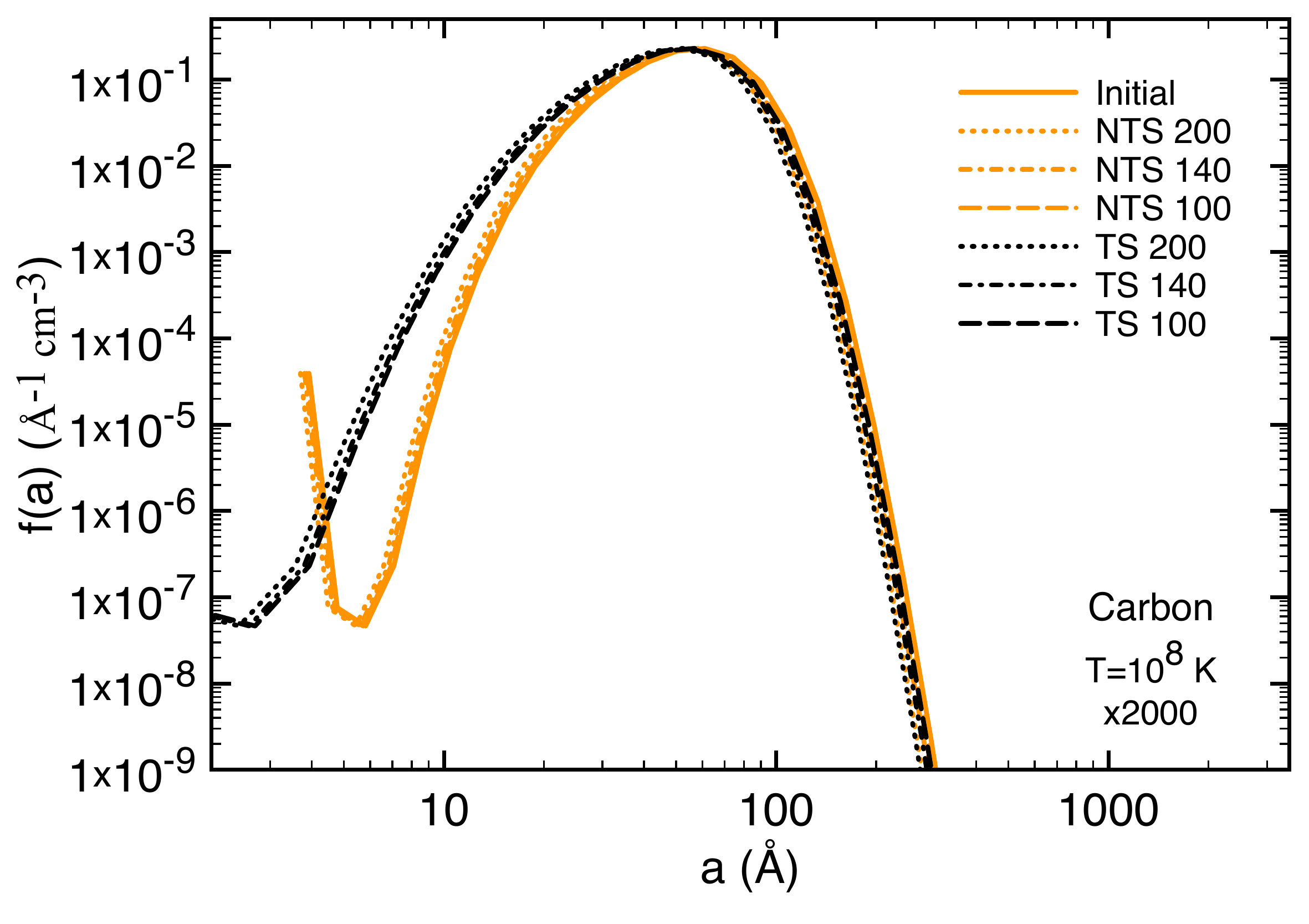}
\includegraphics[width=6cm]{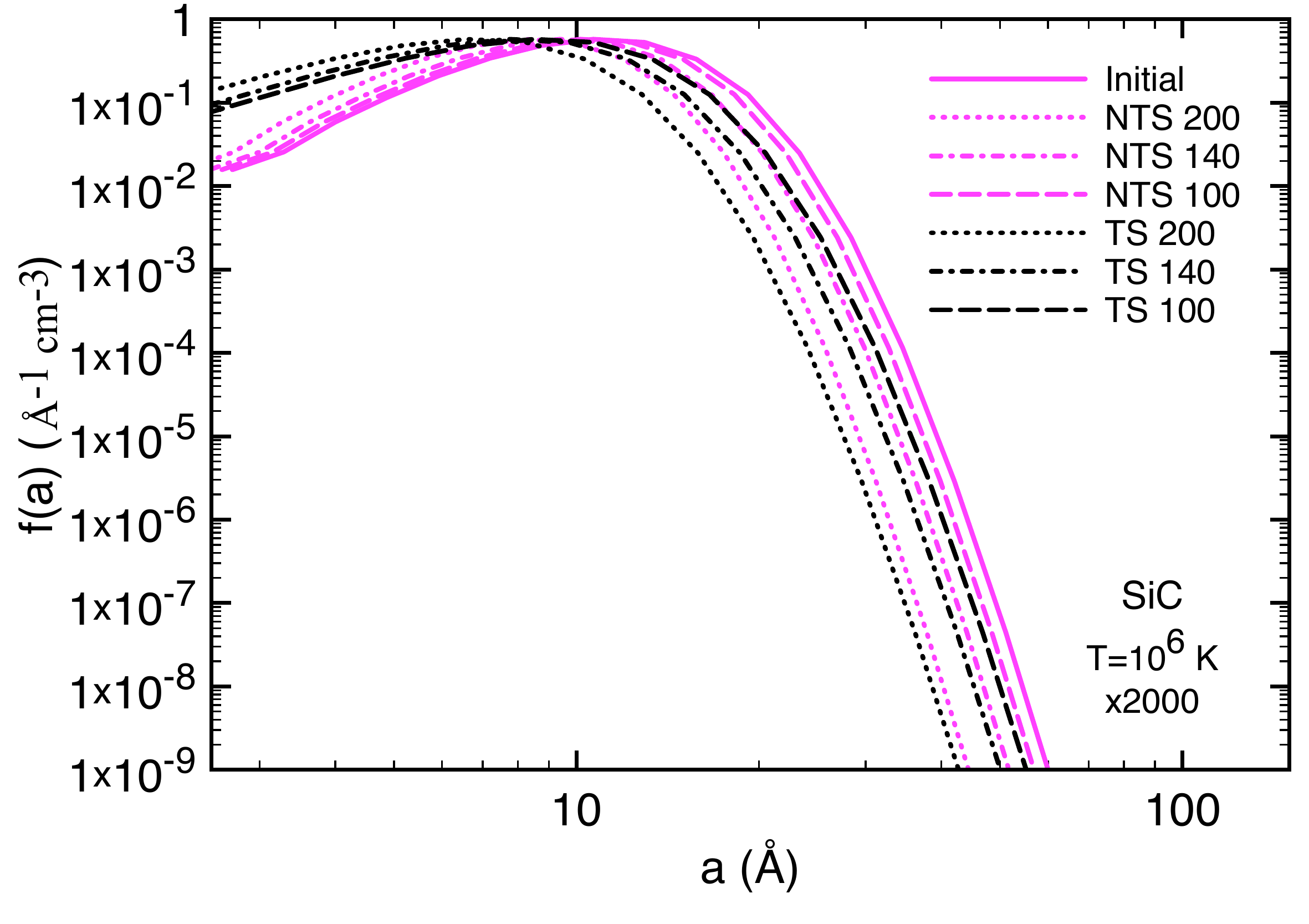}
\includegraphics[width=6cm]{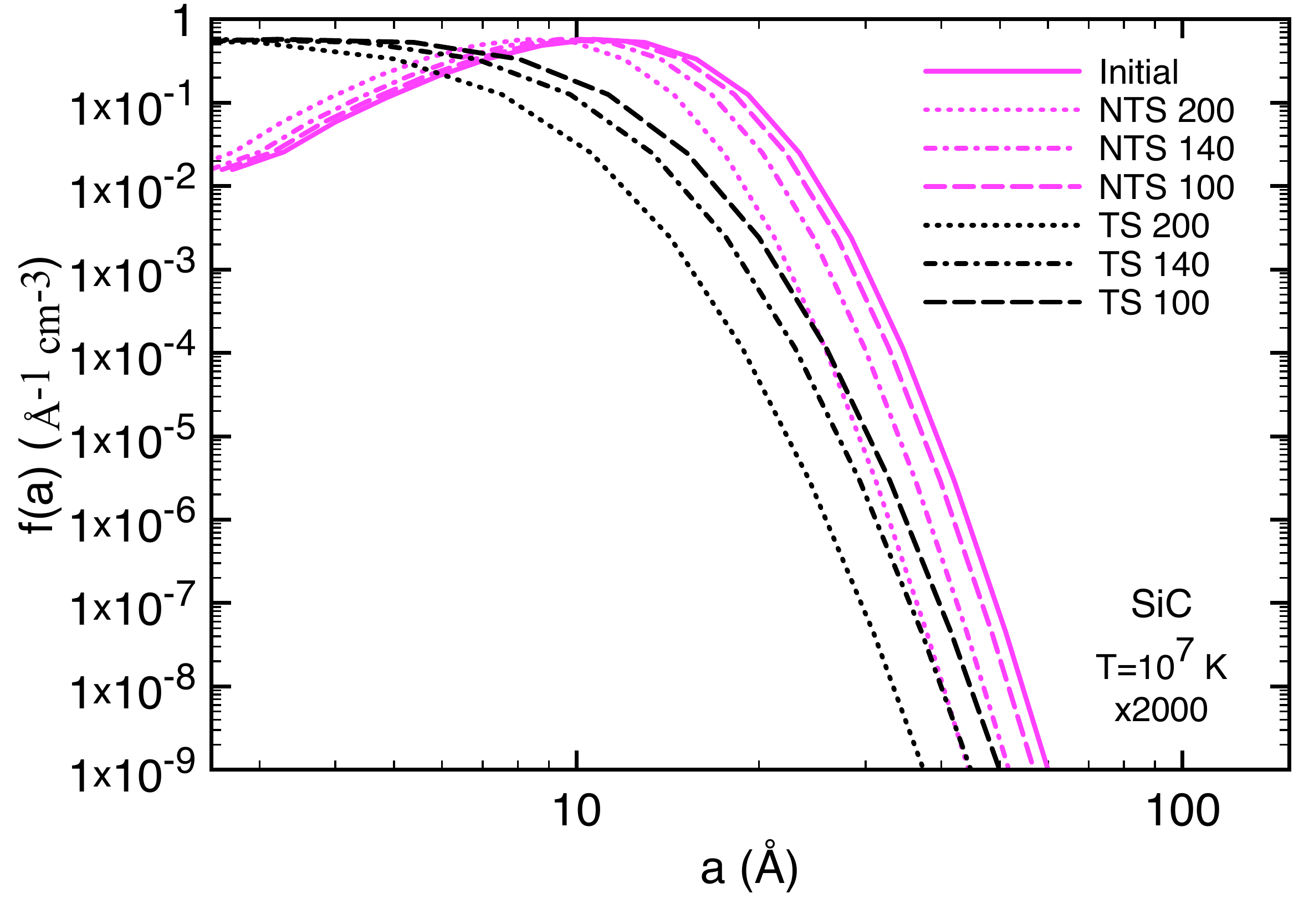}
\includegraphics[width=6cm]{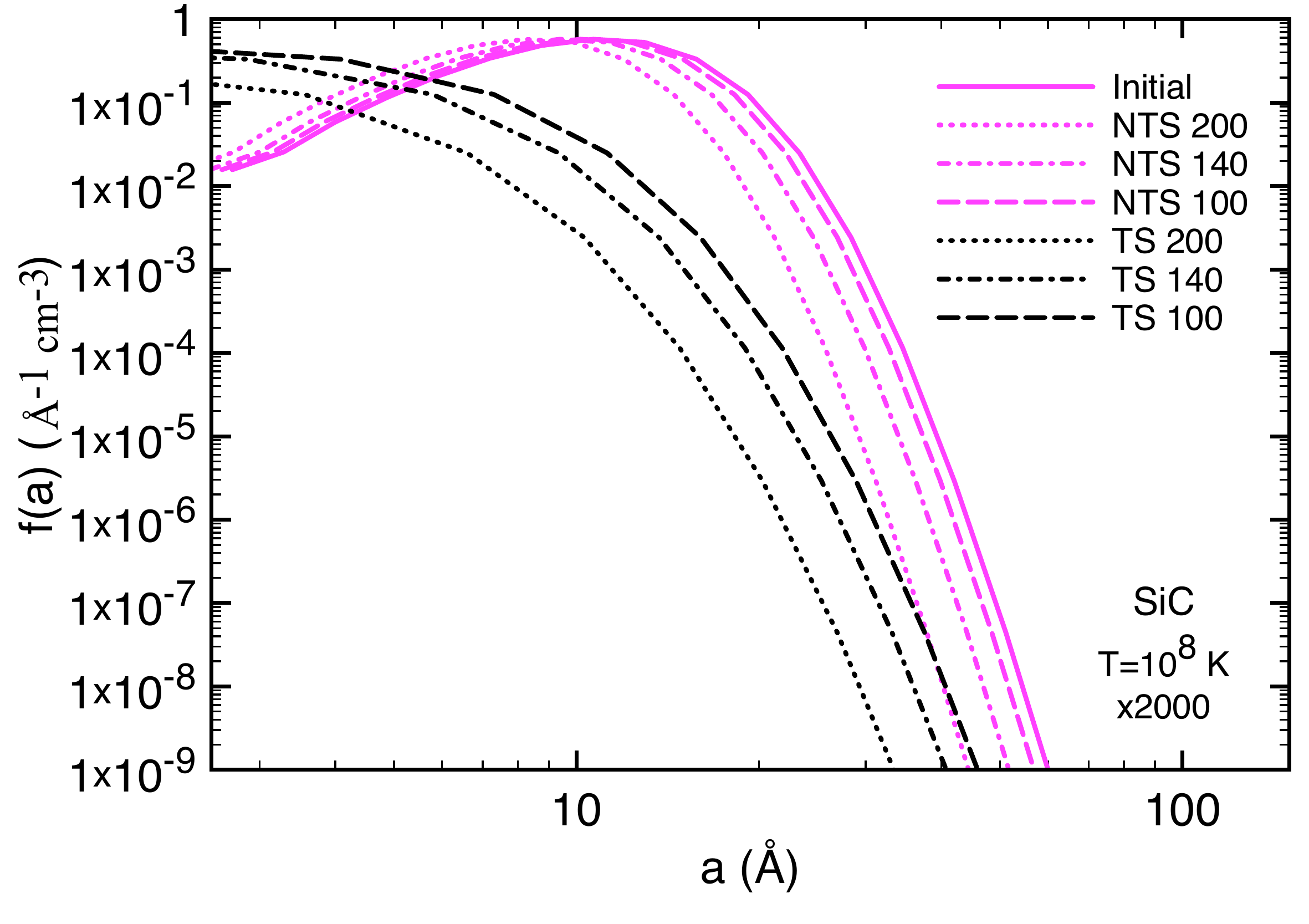}
\caption{Same as Figures \ref{fig5} and \ref{fig6} for the case of a Type~II-P homogeneous ejecta clump (x2 000 case) shocked by the reverse shock. The initial size distributions are plotted with a full line, size distributions after non-thermal sputtering within the clump are plotted with coloured dashed ($v_s=100$ \kms), dashed-dotted ($v_s=140$ \kms), and dotted lines ($v_s=200$ \kms) and after thermal sputtering with black lines for three values of the inter-clump gas temperature ($10^6$ K, $10^7$ K and $10^8$ K).}

\label{fig8}
\end{figure*}

As mentioned before, the x2 000~case corresponds to gas densities typical of Type~II-P~SN homogeneous ejecta (SC15). By homogeneous, we mean that the ejecta is not clumpy and the gas densities and temperature are those provided by explosion models (e.g., Umdeda \& Nomoto 2002). We first discuss homogeneous ejecta and then consider a clumpy ejecta for a Type II-P SN with progenitor mass 19 \Ms, which was chosen as a surrogate for SN1987A in SC15 (although SN1987A is a Peculiar Type II SN). The clumpiness conditions in the ejecta were derived from the modelling of the observed UV, optical, and near IR lines, $\sim$ eight years after explosion (\cite{jerk11}). 

\subsubsection{Homogeneous ejecta}

Results for the grain size distributions are presented in Figure~\ref{fig8} for the four materials of interest. The trends derive for the x200~case still apply, i.e., carbon is by far the most resilient material and survives both non-thermal and thermal sputtering in the clump and the inter-clump medium, respectively. The higher clump density leads to almost no NT sputtering of the carbon grains that were formed in the ejecta, whose size distributions peak at $\sim 60$~\AA. These grains survive thermal sputtering for an inter-clump temperature in the range $10^6-10^8$~K. Therefore, carbon dust formed in Type~II-P ejecta, even of modest size, could survive the RS and the hot inter-clump temperatures.

Silicon carbide is the second most resilient dust type, despite the initial small size of the grains formed in the ejecta. The NT sputtering within the clump is more severe than for carbon since the erosion rate for NT sputtering is higher (see Figure \ref{fig3}) for all shock velocities in the clump. Thermal sputtering is also more effective for SiC than carbon because of the higher erosion rate shown in Figure \ref{fig4}. For the highest inter-clump temperature, the final grain population consists of very small grains or very large SiC clusters with radii less than $10$~\AA, while some grains with~$a > 10$~\AA\ survive the RS for milder inter-clump gas temperatures. 

Dust from the oxygen-rich core follows the same trends as in the x200 case. However, the grains formed in the ejecta being larger (silicate and alumina distributions peak at $\sim 200$~\AA\ and $\sim 150$~\AA, respectively), and the gas density in the clump being higher, the silicate and alumina grains suffer more NT sputtering within the dense clump. However, the final distributions that enter the inter-clump medium at clump disruption all peak at larger radius, and for mild RS velocities and inter-clump gas temperature, the grains can survive thermal sputtering to some extent. For the high inter-clump gas temperatures, thermal sputtering remains severe.  

The dust masses after 4 000 years for the four dust types are summarised in Table \ref{taba1} of the Appendix as a function of RS velocities and inter-clump gas temperatures. The total dust mass values and survival efficiencies are listed in Table \ref{tab4}. We see that NT sputtering within the clump severely destroys dust grains and only $\sim 20-47$~\% of the dust is injected in the inter-clump medium. As expected, these efficiency values are lower than for the x$200$ ~case (see Table \ref{tab3}). For the small grains formed in the SN that led to Cas A, NT sputtering is less severe because the relative velocities in the post-shock gas are small. On the other hand, the large grains that were formed in Type~II-P SNe suffer NT in the dense clump. After thermal sputtering, between $\sim$~14~\% and 45~\% of the dust survives in the remnant, depending on the inter-clump gas temperature. When compared to the results of Table~\ref{tab3}, where $7-11$~\% of the dust survive at $10^7$~K in the x$200$~case, we see that the large dust grains that survive NT sputtering better survive thermal sputtering ($\sim 15-30$~\%) in the x$2 000$ case for a similar inter-clump temperature.   

\begin{table}
\centering      
\caption{Total dust mass (in \Ms) and survival efficiency left after NT and thermal sputtering as a function of RS velocities ($V_{sc}$) and inter-clump temperature (T) for the remnant of a Type~II-P SN (case x2 000). The total initial dust mass before sputtering is $3.2\times~10^{-2}$~\Ms}   
\label{tab4}                      
\begin{tabular}{l c c c c }         
\hline\hline     
\noalign{\vskip 0.7mm}    
 T(K) &  M(NT) & $\epsilon_{NT}$ & M(Thermal) & $\epsilon$  \\
\hline\hline
\multicolumn{2}{c}{ } $V_{sc}=100$ \kms & &  &  \\
\hline
\noalign{\vskip 0.7mm}   
$10^6$&&&$1.4 \times 10^{-2}$&44.6\%\\
$10^7$&$1.5 \times 10^{-2}$&46.8\% &$9.6 \times 10^{-3}$&29.7\%\\
$10^8$&&&$6.2 \times 10^{-3}$&19.3\%\\
\hline\hline
\multicolumn{2}{c}{ } $V_{sc}=140$ \kms & &  &  \\
\hline
\noalign{\vskip 0.7mm}   
 $10^6$ & &&$9.3 \times 10^{-3}$ & 28.8\%  \\ 
 $10^7$ &  $9.7 \times 10^{-3}$ &30.0\% &$6.8 \times 10^{-3}$ & 21.1\% \\  $10^8$ &   & &$5.3 \times 10^{-3}$ & 16.6 \%  \\
\hline\hline
\multicolumn{2}{c}{ } $V_{sc}=200$ \kms & &  &  \\
\hline
\noalign{\vskip 0.7mm}   
 $10^6$ &  &&$6.2 \times 10^{-3}$ & 19.4\%   \\
 $10^7$ &  $6.5 \times 10^{-3}$ &20.3\% &$5.1 \times 10^{-3}$ & 15.8\% \\  $10^8$ &    & &$4.6 \times 10^{-3}$ & 14.3\%   \\
\hline                                    
\end{tabular}
\end{table}

\subsubsection{Clumpy ejecta}
\begin{figure*}
\centering
\includegraphics[width=9.17cm]{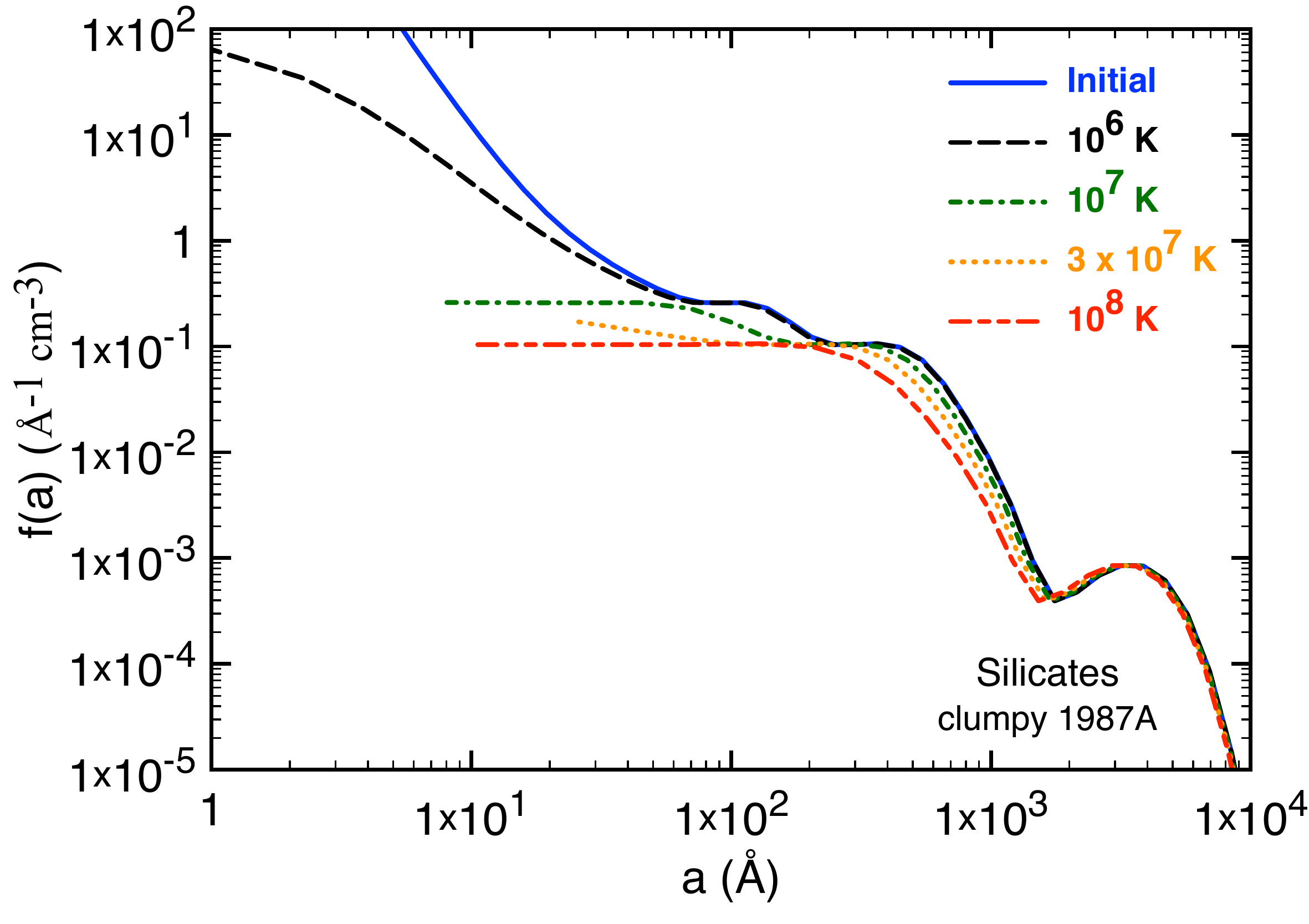}
\includegraphics[width=9.13cm]{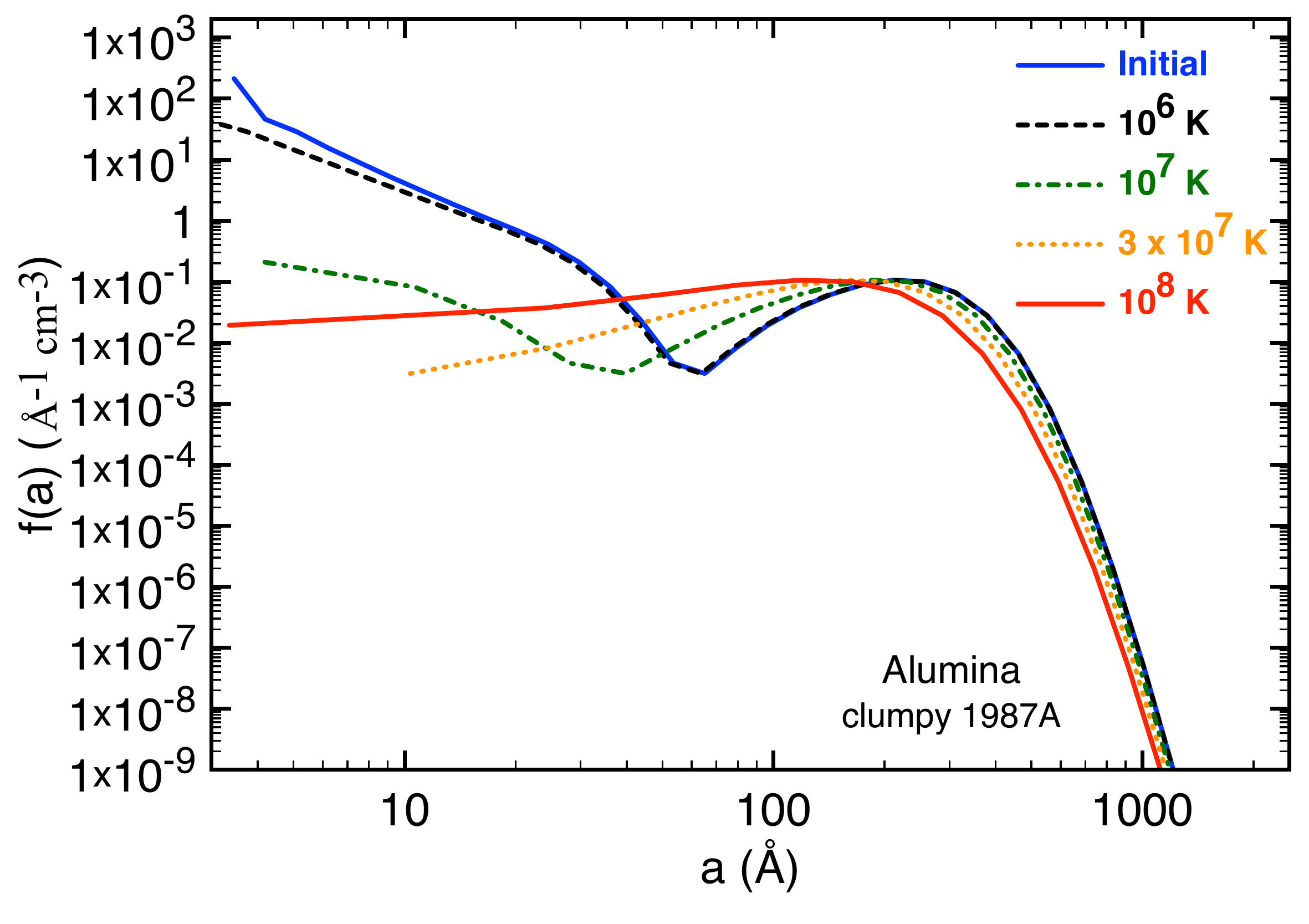}
\includegraphics[width=9.03cm]{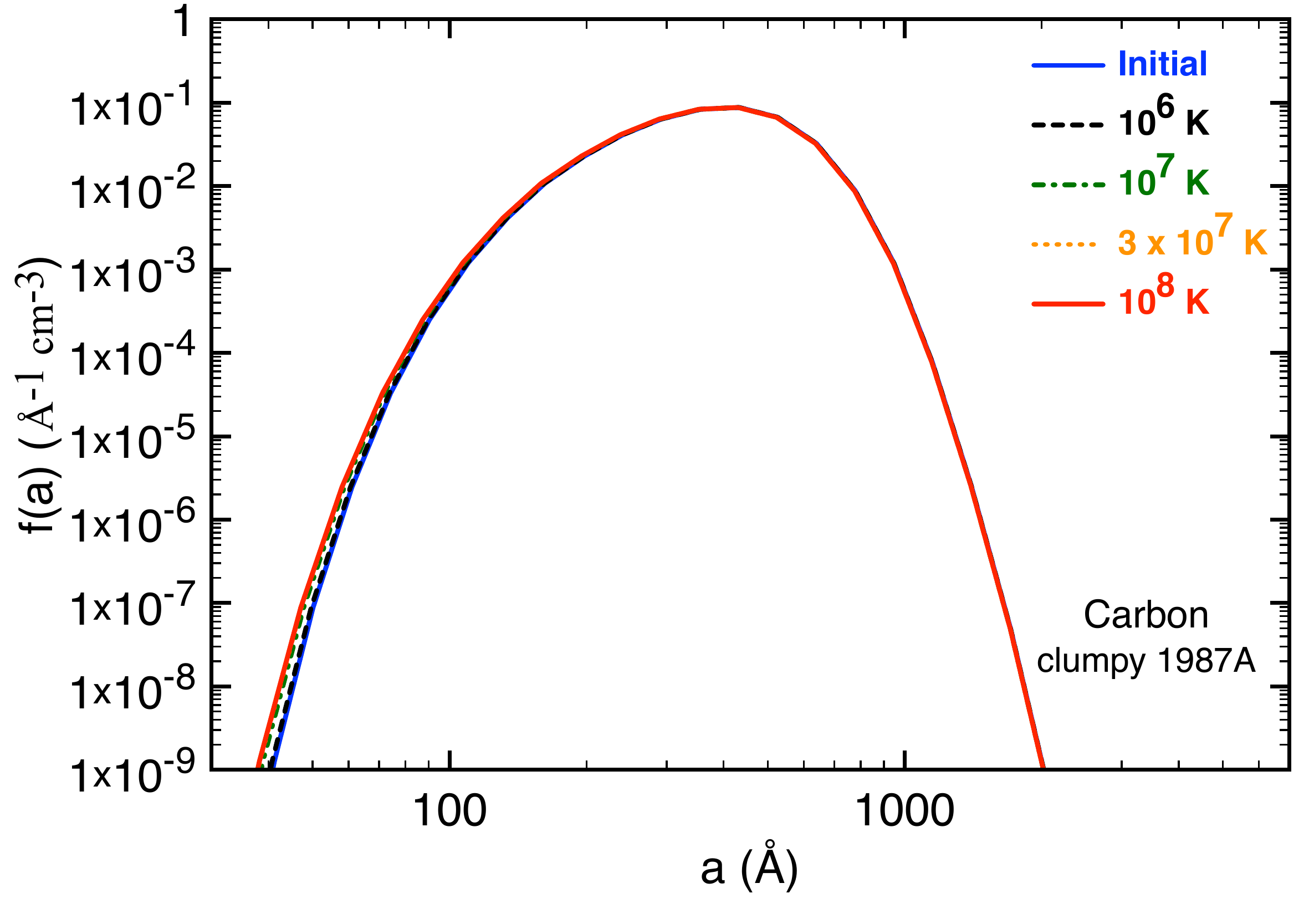}
\includegraphics[width=9.14cm]{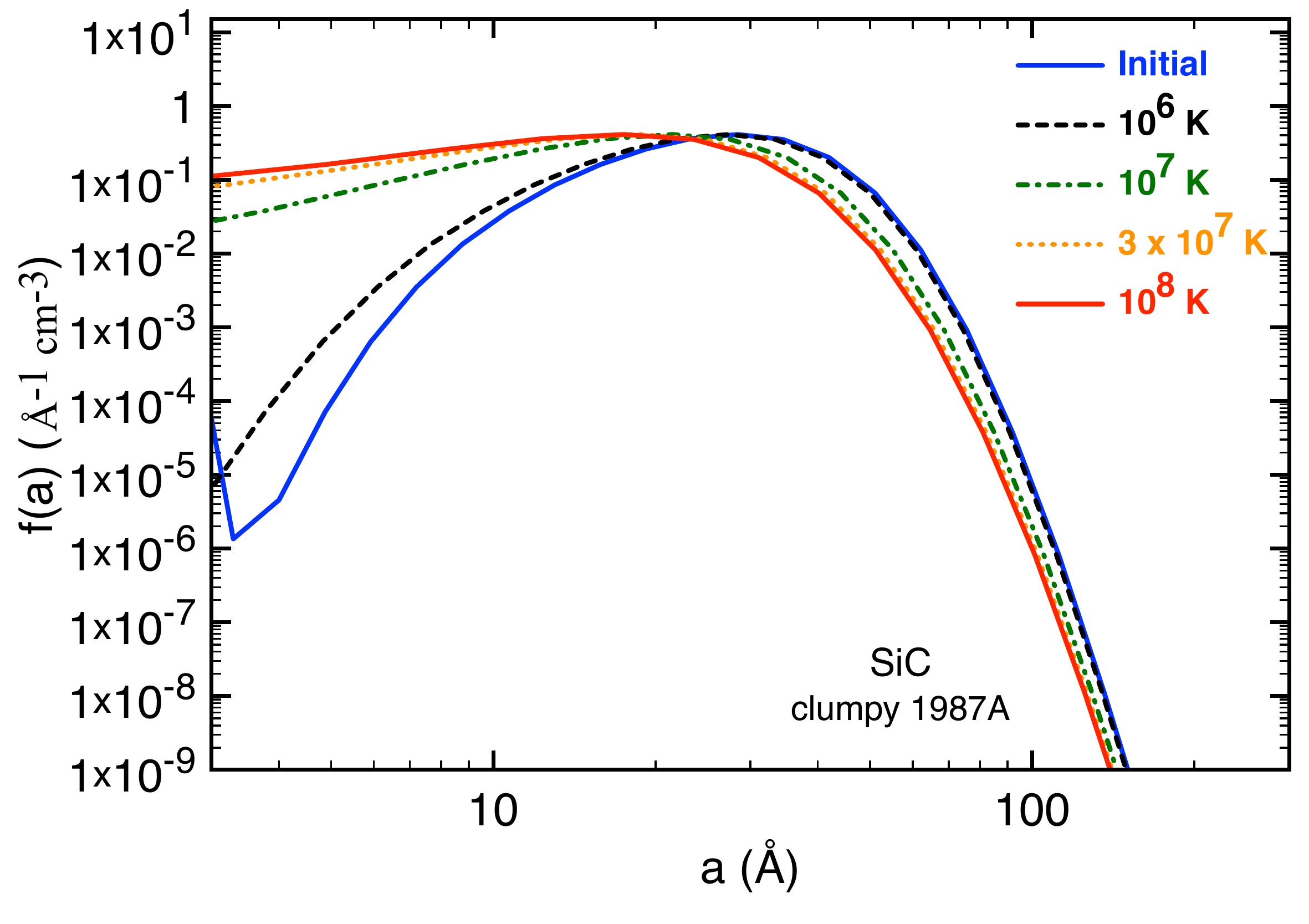}

\caption{Grain size distributions after thermal sputtering in the inter-clump medium as a function of inter-clump temperature for the clumpy Type~II-P ejecta. The initial size distribution for each material is plotted with a full blue line and corresponds to that calculated by Sarangi \& Cherchneff (2015).}

\label{fig10}
\end{figure*}

We now study the sputtering of grains formed in the clumpy ejecta of the 19 \Ms\ progenitor used as a surrogate of SN1987A by SC15. The densities of the various ejecta zones were modelled by assuming the zones were clumpy with over-density factors derived from the modelling of UV, optical and near IR emission lines by Jerkstrand et al. (2011). This clumpy ejecta produces a total dust mass of $0.14$~\Ms\ and the prevalent dust components are alumina, silicate, carbon and iron. For an inter-clump gas number density of 1~\cmc\ at 340 years, the clumps represent inhomogeneities with an over-density factor $\chi \geq1 000$. For such high $\chi$ values, the shocks, initially with velocities in the range $1 000-2 000$~\kms, are decelerate to low velocities in the value range $\sim 10-50$ \kms\ when they cross the clump. At such low values, the relative velocities between the dust grains and the gas within the shocked clump are low and, according to Figure~\ref{fig3}, the NT sputtering rate is negligible for all grain sizes and types. Therefore, we only consider the effect of thermal sputtering in the inter-clump medium for grains formed in the clumpy Type~II-P~SN ejecta. 

Figure \ref{fig10} shows the grain size distributions after 4 000 years of thermal sputtering in the inter-clump medium. For all inter-clump temperatures, the sputtering is minimal. As seen from the initial grain size distributions plotted in Figure \ref{fig10}, the grains in the clump have large sizes before sputtering begins compared to the x$200$ and x$2 000$ cases, and can then better survive thermal sputtering. In particular, all silicate grains with $a \geq 0.15$~\mic\ survive thermal sputtering for all inter-clump temperatures. The final masses of grains after 4 000 years are summarised in Table~\ref{tab5}, where the survival efficiency of the total dust varies from 42~\% to 98~\%. While carbon grains are hardly affected by thermal sputtering at all inter-clump temperatures, silicate and alumina still suffer sputtering at $10^7-10^8$~K, with a survival efficiency range of $32-74$~\%. The composition of the total surviving dust is still dominated by silicate and alumina, but the fraction of carbon dust increases with time and even supersedes that of alumina for the hottest inter-clump gas temperature. 

\begin{table*}
\centering      
\caption{Dust mass (in \Ms) and survival efficiency $\epsilon$ after 4 000 days as a function of dust type and inter-clump gas temperature (T) for the Type~II-P SN clumpy ejecta.}    
\label{tab5}                     
\begin{tabular}{c c c c c c}         
\hline    
\hline
\noalign{\vskip 0.7mm}   
T(K) & Silicate & Alumina & Carbon & SiC & Total dust \\
\hline
\noalign{\vskip 0.7mm}     
Initial&$5.42 \times 10^{-2}$&$1.78 \times 10^{-2}$&$7.49 \times 10^{-3}$&$1.62 \times 10^{-5}$&$7.95 \times 10^{-2}$\\
\hline
\noalign{\vskip 0.7mm}       
10$^6$&5.27 $\times$ 10$^{-2}$&1.75 $\times$ 10$^{-2}$&7.47 $\times$ 10$^{-3}$&1.47 $\times$ 10$^{-5}$& 7.77 $\times$ 10$^{-2}$\\
$\epsilon$& 97.3\%\ &98.2 \% & 99.6 \% & 90.7 \%& 97.6 \% \\
\hline
\noalign{\vskip 0.7mm}   
10$^7$&4.01 $\times$ 10$^{-2}$&1.33 $\times$ 10$^{-2}$&7.38 $\times$ 10$^{-3}$&8.43 $\times$ 10$^{-6}$&6.07 $\times$ 10$^{-2}$\\
$\epsilon$& 73.9\%\ &74.5\% & 98.5\% & 52.0\% & 76.3\% \\
\hline
\noalign{\vskip 0.7mm}    
10$^8$&2.07 $\times$ 10$^{-2}$&5.68 $\times$ 10$^{-3}$&7.35 $\times$ 10$^{-3}$&5.52 $\times$ 10$^{-6}$& 3.37 $\times$ 10$^{-2}$\\
$\epsilon$& 38.2\%&32.0\%& 98.1\% &34.1\% & 42.4\% \\   
\hline
\end{tabular}
\end{table*}

\subsection{Comparison with existing studies}
Several studies on the formation of dust in SNe and its processing in the remnant phase were undertaken by various groups for local or high redshift objects. Despite the fact that the formalism, approach and models are different from those being used in the present study, it is useful to compare trends and test our findings against other studies. 

In their study of the formation and evolution of dust in Type~II-P and Type~II-b SNe, with application to Cas A, Nozawa et al. (2007, 2010) consider the very small grains that formed out of CNT in the diffuse, stratified ejecta of the SN, and their destruction by NT and thermal sputtering in the remnant phase once the RS encounters the ejecta. The ejecta is considered homogenous, i.e., clump-free. For Cas A, a large mass of dust forms ($0.167$~\Ms) for the diffuse conditions of the SN ejecta, owing to the use of CNT. The gas-phase chemistry is overlooked and a large fraction of elements is locked up in molecules. Molecular species indeed represent between $\sim 30$~\% and $50$~\% of the total ejecta mass and regulate the elements available for dust nucleation and condensation (\cite{sar13}). The recent ALMA detection of a CO mass in excess of $0.01$~\Ms, of SiO, SO and HCO$^+$ in SN1987A supports this point (\cite{kam13, mat15b}). Thus, the masses of dust formed by Type~II~SNe are likely to be lower than those predicted by CNT. The shocked gas comprised between the RS and the forward shock is hot and diffuse ($10^6-10^8$ K). Their final conclusion is that grains of any chemical type and with size $a < 0.1$ \mic\ are destroyed by thermal sputtering over a timescale of several thousands of years, being trapped in the hot diffuse gas. Conversely, larger grains with $a > 0.1$ \mic\ can escape the hot, diffuse region and cross the forward shock because they suffer less drag in the diffuse gas, and maintain high relative velocities with respect to the gas. 

Another study of grain formation in SNe and sputtering in Type~II-P SN remnants is provided by Bianchi \& Schneider (2007) and used in subsequent studies (e.g., Chiaki et al. 2015, Marassi et al. 2015). They consider the formation of dust in fully-mixed ejecta for various SN progenitor masses by also using a CNT formalism. The subsequent NT and thermal sputtering in the post-RS gas is studied for several thousands of years, but occurs in an homogeneous, diffuse ejecta. Their results indicate a mass fraction of $\sim 7-10$~\% of surviving dust for a $18-20$~\Ms\ SN progenitor appropriate for Cas A. However, it is worth pointing out a few drawbacks with the initial model and subsequent studies. Firstly, an SN ejecta is not fully-mixed but shows some stratification in its chemical composition, as indicated by atomic line emission in the remnant phase (\cite{ise12}). The SN models with $18-20$~\Ms\ progenitors that were considered by Bianchi \& Schneider produce a large dust mass ($\sim 0.2-0.4$ \Ms) because the ejecta is fully-mixed and CNT is used to model dust formation, for which the above caveats apply. Secondly, the derived dust composition is dominated by large carbon grains in most models. SNe with progenitor masses of 18 \Ms\ or above primarily produce silicates and oxides with a small component of carbon dust (Cherchneff \& Dwek 2010; SC15). Emission lines of SiO are detected in some relatively massive Type~II-P SNe and correlate with the formation of silicate dust, e.g., in SN2004 (\cite{ko09}). Furthermore, amongst the  dust assumed to form in their models, magnetite, Fe$_3$O$_4$, has so far escaped detection in pre-solar grains of SN origin while alumina oxides have been identified and seem to be ubiquitous (\cite{hop11}). Therefore, it is likely that the dust composition formed in a $18-20$ \Ms\ progenitor SN includes a prevalent component of silicates and oxides, as opposed to carbon dust. 

Dynamical studies of dust thermal sputtering in clumps disrupted by the RS are proposed by Silvia et al. (2010, 2012). These studies model the interaction of the RS with a spherical clump in which a population of dust grains with initial size distributions is implanted. Two values of the density contrast parameter $\chi$ are considered (100, 1 000) and various RS velocities are investigated. The initial dust mass and size distributions are those calculated by Nozawa et al. (2003), and simulations are run over a period of $10\times \tau_{cc}$, with a maximum time of 1 000 years. Overall, the RS velocities and $\chi$~factors considered in these studies are similar to those that we consider in the present work. However, thermal sputtering is assessed within the clump affected and finally disrupted by the RS for gas temperatures that do not exceed $10^6$~K. Despite the difference in approach and model, the survival efficiencies for carbon dust are consistent with our findings. Conversely, the results for alumina and silicate dust differ, partly because of the difference in initial dust size distributions. Indeed, the alumina grains produced by Nozawa et al. (2003) from CNT-based models have very small sizes ($a < 40$ \AA) while the alumina distribution in Figure \ref{fig1} peaks at  $\sim 150$~\AA. No alumina grains survive the RS for all cases investigated by Silvia et al. On the other hand, silicate dust better survives the RS in their simulations because the silicate grain size distribution derived from CNT extends to large radii ($100$~\AA$< a< 1 000$~\AA), while our silicate grain size distribution in Figure \ref{fig1} peaks at $\sim 150$~\AA.  

Finally, Micelotta \& Dwek (2013) explore the destruction of dust in Cas A with an approach similar to our clump model. They only consider one $\chi$ factor value (100) and one inter-clump temperature ($10^8$~K). They use ad-hoc Mathis-Rumpl-Nordsieck (MRN) size distributions (\cite{mat77}) for carbon and silicate dust formed in the Cas~A SN progenitor and do not consider other dust types, e.g., alumina. A MRN grain size distribution is appropriate for ISM dust and may not apply to SN dust. Indeed, SC15 show that the size distributions of grains of various dust types, which are produced in the ejecta of Type~II~SNe 2~000~days after the explosion, strongly deviate from a MRN distribution, are skewed towards large grain sizes and contain grains with radii below 50~\AA. Because the sputtering results are dependent on the initial dust type and size, the application of a MRN distribution to two dust components may not adequately reflect the situation of dust formation and processing in the Cas~A remnant. 
\section{Summary and discussion}
\label{dis}
We have modelled the sputtering that is experienced by dust grains that formed in the ejecta of Type~II-b and Type~II-P SNe in the remnant phase. We assume the ejecta clumps are crossed by the reverse shock and consider both non-thermal sputtering within the clumps and thermal sputtering in the inter-clump medium, for various density contrasts, shock velocities and inter-clump temperatures. The main results are as follows:

\begin{itemize}
\item For Cas  A, our model shows that between 30\% and 60 \% of the $\sim 0.03$ \Ms\ of dust formed in the ejecta is present today. Only 6~\% to 11~\% of the initial mass will survive the remnant phase after 4 000 years for all RS velocities and the two inter-clump temperatures that characterise the remnant. The prevalent surviving dust is carbon with some trace of SiC and alumina while silicate grains will be totally destroyed. 
\item Our modelled dust masses that are currently being sputtered in Cas~A (between 30~\% and 60~\% of the initial mass) agree well with values derived from mid-IR observations of the remnant. The results indicate that the dust in Cas A is currently eroded by NT sputtering within the ejecta dense clumps.
\item For Type~II-P SNe, the initial grain size distributions that peak at larger radii, larger mass fractions of dust survive with efficiency values in the range $14-45$~\%.
 \item For the clumpy SN with a 19~\Ms\ progenitor, which is a surrogate of SN1987A, surviving mass fractions are between 42~\%\ to 98~\%, where the composition of the surviving dust is made of alumina, silicates, carbon, and a small fraction of SiC. According to SC15, pure metal dust, namely iron, silicon and magnesium, could also be present, but have not been considered in this study. If we add up these dust components to the initial dust mass, between $\sim 0.06$~\Ms\ and $0.13$~\Ms\ of dust with grain size in excess of $\sim 0.04$~\mic\ can survive up to 4 000 years of sputtering in the remnant phase. 
\item{For Cas~A and in Type~II-P SNe, non-thermal sputtering within the clumps is severe for all grains and reduce the surviving dust mass to the value range $28-58$~\% and $20-47$~\%, respectively, before thermal sputtering intitiates. The oxygen-rich dust grains, namely silicate and alumina, are the grains most severely
eroded by NT sputtering in clumps because of their initial large sizes. }
\end{itemize}

The sputtering results are highly sensitive to the initial grain size distributions and the assumed or derived dust chemical composition. This is clear from the present study but was already mentioned by previous studies (\cite{noz07,bian07,sil10}). An illustrative example is provided by alumina grains, for which Nozawa et al. (2006) derived small radii ($a \leq 40$ ~\AA). These size distributions were later used by Silvia et al. (2010, 2012) as initial conditions to their sputtering analysis. In both studies, alumina grains are totally destroyed after several hundreds of years, which is in contrast to the findings of alumina pre-solar grains of SN origin in meteorites, which represent about 10 \% of the whole pre-solar alumina grain budget (\cite{hop11}). These studies model dust formation in SN ejecta by using the CNT formalism. As already mentioned, CNT assumes gas equilibrium conditions, which are inappropriate for the description of a shocked gas phase in an environment controlled by radioactivity (\cite{do85,cher09}). The approach does not discriminate between possible dust chemical compositions because, once the condensation temperature at equilibrium for a given material is reached, CNT assumes dust can form. This results in a large variety of possible condensates (e.g., magnetite Fe$_2$O$_3$, as in Bianchi \& Schneider 2007) that have not been identified in mid-IR SN emission spectra or in pre-solar grains. A chemical kinetic approach is preferred to model dust formation because it describes the formation of molecules, dust clusters, and grains under realistic, non-equilibrium gas conditions, and considers the crucial step of dust nucleation, which is controlled by the external physics of the environment (e.g., radioactivity). These models produce dust compositions and size distributions that are different from those derived from CNT, and kinetic models usually produce lower dust masses made of fewer dust materials than those based on CNT. In view of the sensitivity of sputtering to initial dust grain sizes and chemical compositions, it is important to adopt a complete approach of dust formation in SNe that describes both the gas phase (molecules) and the solid phase through formation of dust clusters and their condensation. We note that chemical kinetic models (e.g., Cherchneff \& Dwek 2009, 2010; Sarangi \& Cherchneff 2013, 2015; BC14) have so far made predictions on molecules and dust types in SNe and SNRs that have been supported by Herschel and ALMA data (\cite{kam13,wal13,mat15b}). 

Dust formation and sputtering are also very sensitive to the level of clumpiness of the SN ejecta and the remnant phase. In the case of Type~{II-b}~SNe with Cas~A in mind, BC14 show that density increments of several hundreds with respect to homogeneous ejecta conditions are necessary to explain the diversity of dust grains that formed in Cas~A, as revealed by mid-IR spectra analysis (\cite{dou01,rho08,ar14}). Indeed, no dust clusters and grains can form for the diffuse conditions of ejecta models given by Nozawa et al. (2010). To illustrate the above caveats regarding CNT, Nozawa et al. find that $0.167$~\Ms\ of dust forms in Cas~A by using CNT for similar diffuse gas conditions. Clumping is required to permit the formation of large dust grains after an SN explosion. Such large grains can then endure the NT and thermal sputtering in the remnant phase. For silicates, large grains with $a \geq 0.1$~\mic\ only form in the clumpy SN surrogate of SN1987A (SC15). Therefore, silicate pre-solar grains probably originate from such dense, clumpy Type~II-P SN explosions. A similar conclusion holds for alumina grains, which need to form in dense ejecta clumps with large enough radii ($a \geq 0.05$~\mic) to survive sputtering in the remnant phase and explain the origin of pre-solar alumina grains. A recent investigation by Nozawa et al. (2015) reaches a similar conclusion regarding the necessity of clumps in SNe and their remnants to explain pre-solar alumina material, although the analysis and results differ from those of BC14 and of the present model owing to the use of CNT and of a homogeneous remnant phase. 

Clumps are also essential to the formation of large carbon-rich dust grains in SNe. We see from Figure \ref{fig10} that the grains of SiC that form in our clumpy SN1987A surrogate remain very small, with most of the grains having radii $a \leq 70$~\AA. Carbon and SiC grains form in the outermost zone of SN ejecta, which is pervaded by He$^+$ ions that hamper the formation of molecules and dust clusters. Molecules and dust clusters can form only when He$^+$ stops forming or recombine (\cite{sar13}). Models usually assume stratified ejecta in which there is no leakage between zones, i.e., the carbon-rich zone is controlled by He, carbon, and oxygen.  However, this zone confines with the hydrogen-rich progenitor envelope. SN explosion models based on 3D hydrodynamics indicate hydrogen penetrates the carbon-rich clumps and microscopically mixes with the gas (\cite{wo15}). Therefore, the current hydrogen-free chemistry might be inappropriate to describe carbon and SiC dust synthesis for a high level of hydrogen mixing. The inclusion of hydrogen will trigger a reactive radical chemistry, will probably alter results on the local chemistry of SiC dust, and will hopefully help producing large SiC grains such as the $\sim 1$~\mic-size, Type X SiC pre-solar grains found in meteorites (\cite{hyn06}). 

The fate of dust grains in SN remnants also depends on clumpiness and is altered by considering an ejecta made of clumps and knots. Existing studies (Nozawa et al. 2007, 2010, 2015; Bianchi \& Schneider 2007) consider a homogeneous ejecta gas that is crossed by the RS. In the post-shock gas, the large grains will have high relative velocities with respect to the gas in the immediate post-shock environment. Following Nozawa et al., the grains will travel across the diffuse region between the RS and the forward shock while experiencing NT and thermal sputtering. Because the medium is diffuse, the large grains maintain high relative velocities despite sputtering, are able to cross the forward shock and reach the interstellar medium. In the present study, large grains experience NT sputtering within the dense clumps and will gradually decelerate as the relative velocity drops. Therefore, large grains will be trapped in the clump until the clump is disrupted by the crossing of the RS. They will then experience thermal sputtering in the inter-clump medium and will expand with the gas until they reach the denser gas at the back of the forward shock. However, large grains will not cross the forward shock because of their small relative velocities. This scenario is in contrast to conclusions by Nozawa et al. and requires a means to carry large grains across the forward shock. A solution to this quandary may reside in the outcomes of hydrodynamic simulations of shock-cloud interactions. Indeed, Silvia et al. (2010) find that the RS disrupts clumps and create small, dense regions, or bullets, when radiative cooling by metals is considered. Such structures are predicted by other 3D hydrodynamic studies of shock-cloud interaction (e.g., Pittard \& Parkin 2015) and small, dense, optically emitting knots embedded in an X-ray emitting, diffuse gas are observed in Cas~A (\cite{pat14}). The 3D kinematic reconstructions of the optical emission of few SN remnants, e.g., Cas~A (\cite{mil13}) and SN~1E0102.2-7219 in the Small Magellanic Cloud (\cite{vog10}), show the ejecta shocked by the RS are characterised by knots, rings, an overall velocity asymmetry in the bulk material and the presence of fast, clumpy jets that may relate to the explosion dynamics. Such complex remnant structures and dynamics are in marked contrast with the remnant homogeneous ejecta that are assumed in several dust-sputtering studies. We may then speculate that, at late evolutionary time, the large dust grains that are trapped in these dense bullets will find their way to the interstellar medium, as the bullets may retain a memory of the original knot or jet dynamics. Based on the results for Cas~A and the clumpy SN1987A case, for which a total dust mass in the range $0.06-0.13$ \Ms\ survives to sputtering in the remnant phase, we conclude that Type~II-b SNe may not provide much dust to the interstellar medium, while SNe with clumpy, dense ejecta and their remnant could significantly contribute to the dust budget of local and high redshift galaxies.  

\acknowledgements
The authors thank the anonymous referee for constructive comments and Arkaprabha Sarangi for supplying the dust grain size distributions and for constructive discussions. C.B. acknowledges support from the Swiss National Science Foundation grants 20GN21-128950, 200020-149248 and P2BSP2-158777. I.C. acknowledges support from the European Research Council (FP7) under the ERC Advanced Grant Agreement No. 321263 - FISH.

\appendix
\section{}
\begin{table*}    
\caption{Dust masses after 4 000 days as a function of dust type, RS velocities in the clump, $V_{sc}$, and inter-clump gas temperatures for the homogenous (x2 000) case. The survival efficiency $\epsilon$ is listed, as well as the survival efficiency after NT sputtering in the clump, $\epsilon_{NT}$.  }   
\label{taba1}  
\centering                      
\begin{tabular}{c c  c c  c l }         
\hline    
\hline     
&T(K) & Initial & NT & Thermal & $\epsilon$ (\%) \\
\hline
\hline
\multicolumn{2}{l}{$v_{sc}  =100$ \kms}  & &  & & \\
\hline
&$10^6$&   &   & $1.0 \times 10^{-3}$ & 8.9  \\
Silicate&$10^7$ &$1.1 \times 10^{-2}$ & $1.3 \times 10^{-3}$  &  $4.8 \times 10^{-5}$ & 0.4 \\
&$10^8$&  & $\epsilon_{NT} =11.8 $\%&  $1.6 \times 10^{-8}$ &  0.   \\
\hline
&$10^6$&   &  &  $6.7 \times 10^{-3}$ & 47.6  \\
Alumina&$10^7$ &$1.4 \times 10^{-2}$ & $7.0 \times 10^{-3}$ & $3.4\times 10^{-3}$ & 24.1 \\
&$10^8$&  &  $\epsilon_{NT} =50.0 $\%& $3.0 \times 10^{-4}$  &  2.1  \\\hline
 & $10^6$&   &   & $6.6 \times 10^{-3}$ & 92.1  \\
Carbon&$10^7$ &$7.1 \times 10^{-3}$ & $6.7 \times 10^{-3}$ & $6.1\times 10^{-3}$ & 85.0  \\
&$10^8$&  &  $\epsilon_{NT} =94.4 $\%& $5.9 \times 10^{-3}$  & 82.5  \\\hline
 &$10^6$&   &   & $6.0 \times 10^{-5}$ & 65.9  \\
SiC&$10^7$ &$9.1 \times 10^{-5}$ & $7.9 \times 10^{-5}$ & $1.0\times 10^{-5}$ & 11.5 \\
&$10^8$&  &  $\epsilon_{NT} =86.8 $\%& $2.2 \times 10^{-6}$  & 2.4 \\
\hline
\hline   
\multicolumn{2}{l}{$v_{sc}  =140$ \kms} & &  & &  \\
\hline
&$10^6$&   &   & $1.1 \times 10^{-4}$ & 1.0 \\
Silicate &$10^7$ &$1.1 \times 10^{-2}$ & $2.0 \times 10^{-4}$ & $1.8 \times 10^{-7}$ & 0. \\
&$10^8$&  & $\epsilon_{NT} =1.8 $ \% & $3.6 \times 10^{-13}$ &  0.  \\
\hline
& $10^6$&   &   & $3.1 \times 10^{-3}$ & 22.4 \\
Alumina&$10^7$ &$1.4 \times 10^{-2}$ & $3.3 \times 10^{-3}$ & $1.3\times 10^{-3}$ & 9.1 \\
&$10^8$&  &$\epsilon_{NT} =23.6 $\% & $3.9 \times 10^{-5}$  &  0.3 \\
\hline
& $10^6$&   &   & $6.0 \times 10^{-3}$ & 84.3 \\
Carbon& $10^7$ &$7.1 \times 10^{-3}$ & $6.1 \times 10^{-3}$ & $5.5\times 10^{-3}$ & 77.6\\
&$10^8$&  &$\epsilon_{NT} =85.9 $\% & $5.3 \times 10^{-3}$  & 75.0  \\
\hline
&$10^6$&   &   & $4.6 \times 10^{-5}$ & 50.3  \\
SiC &$10^7$ &$9.1 \times 10^{-5}$ & $6.1 \times 10^{-5}$ & $6.4\times 10^{-6}$ & 7.0 \\
&$10^8$&  &$\epsilon_{NT} =67.0 $\% & $1.1 \times 10^{-6}$  & 1.2  \\
\hline   
\hline   
\multicolumn{2}{l}{$v_{sc}  =200$ \kms}  & &  & &  \\
\hline
&$10^6$&   &   & $9.8 \times 10^{-6}$ & 0.09 \\
Silicate &$10^7$ &$1.1 \times 10^{-2}$ & $1.2 \times 10^{-5}$ & $1.7 \times 10^{-12}$ & 0. \\
& $10^8$&  &$\epsilon_{NT} =0.1 $\% & $2.7 \times 10^{-22}$ &  0. \\
\hline
&$10^6$&   &   & $1.0 \times 10^{-3}$ & 7.3  \\
Alumina &$10^7$ &$1.4 \times 10^{-2}$ & $1.1 \times 10^{-3}$ & $2.8\times 10^{-4}$ & 1.8\\
& $10^8$&  &$\epsilon_{NT} =7.9 $\% & $6.6 \times 10^{-7}$  &  0.  \\
\hline
&$10^6$&   &   & $5.2 \times 10^{-3}$ & 73.6 \\
Carbon &$10^7$ &$7.1 \times 10^{-3}$ & $5.4 \times 10^{-3}$ & $4.8\times 10^{-3}$ & 67.5\\
&$10^8$&  &$\epsilon_{NT} =76.1 $\% & $4.6 \times 10^{-3}$  & 65.4 \\
\hline
&$10^6$&   &   & $2.9 \times 10^{-5}$ & 31.8\\
SiC&$10^7$ &$9.1 \times 10^{-5}$ & $4.6 \times 10^{-5}$ & $2.8\times 10^{-6}$ & 3.0  \\
&$10^8$&  &$\epsilon_{NT} =50.5 $\% & $3.0 \times 10^{-7}$  & 0.3  \\
\hline                                                        
\end{tabular}
\end{table*}



\end{document}